\documentclass[10pt,oneside,twocolumn,aps,pra,preprintnumbers,bibnotes10pt,superscriptaddress,amsmath,amssymb]{revtex4-2}

\usepackage{graphicx}
\usepackage{float}
\usepackage{braket}
\usepackage{bbold}
\usepackage{amsmath}
\usepackage{amsthm}
\usepackage[dvipsnames]{xcolor}
\usepackage{multirow}
\usepackage{tikz}
\usetikzlibrary{positioning}
\usetikzlibrary{decorations.pathmorphing, decorations.pathreplacing}
\usepackage{ragged2e}
\usepackage{commath}
\usepackage{comment}
\usepackage{adjustbox}
\usepackage{subfig}
\usepackage{hyperref}
\usepackage[normalem]{ulem}
\usepackage{figchild}

\newcommand{\tr}{\text{Tr}}
\newcommand{\id}{\mathbb{1}}

\newcommand{\ketbra}[2]{\ket{#1} \hspace{-1mm} \bra{#2}}
\newcommand{\var}{\Delta^2}

\newcommand{\idmix}{\hat{\rho}_{\id \text{-m}}}
\newcommand{\thmix}{\hat{\rho}_{\text{th-m}}}
\newcommand{\thermal}{\hat{\rho}_{\text{th}}}
\newcommand{\squeez}{\hat{\rho}_{\text{sq}}}

\newcommand{\squeezvec}[1]{\ket{\psi_{\rm sq} (#1)}}

\newcommand{\kghz}{\hat{\rho}_{\text{k-s}}}
\newcommand{\plusghzvec}[1]{\ket{\psi_{\text{+-s}} (#1)}}
\newcommand{\plusghz}{\hat{\rho}_{\text{+-s}}}
\newcommand{\nc}{{\: \text{no-c}}}

\newcommand{\new}[1]{\textcolor{blue}{#1}}

\definecolor{lightgray1}{gray}{0.8}  
\definecolor{lightgray2}{gray}{0.6}  

\usepackage[dvipsnames]{xcolor}

\begin{document}

\title{Nonequilibrium thermometry via an ensemble of initially correlated qubits}

\author{Enrico Trombetti}
\affiliation{Istituto Nazionale di Ottica del Consiglio Nazionale delle Ricerche (CNR-INO), Largo Enrico Fermi 6, I-50125 Firenze, Italy.}

\author{Marco Malitesta}
\affiliation{Istituto Nazionale di Ottica del Consiglio Nazionale delle Ricerche (CNR-INO), Largo Enrico Fermi 6, I-50125 Firenze, Italy.}
\affiliation{Department of Physics and Astronomy, Università di Firenze, Via Sansone 1 I-50019 Sesto Fiorentino, Italy.}

\author{Marco Pezzutto}
\affiliation{Istituto Nazionale di Ottica del Consiglio Nazionale delle Ricerche (CNR-INO), Largo Enrico Fermi 6, I-50125 Firenze, Italy.}

\author{Stefano Gherardini}
\affiliation{Istituto Nazionale di Ottica del Consiglio Nazionale delle Ricerche (CNR-INO), Largo Enrico Fermi 6, I-50125 Firenze, Italy.}
\affiliation{European Laboratory for Non-linear Spectroscopy, Università di Firenze, I-50019 Sesto Fiorentino, Italy.}

\date{\today}

\begin{abstract}
We investigate a nonequilibrium quantum thermometry protocol in which an ensemble of qubits, acting as temperature probes, is weakly coupled to a macroscopic thermal bath. The temperature of the bath, the parameter of interest, is encoded in the dissipator of a Markovian thermalization process. For some relevant initial states, we observe a peak in the Quantum Fisher Information (QFI) during the transient of the thermalization, indicating enhanced sensitivity in early-time dynamics. This effect becomes more pronounced at higher bath temperatures and is further enhanced when the initial reduced state of the qubits has a large ground-state population and/or it is highly coherent. We also focus on the role of initial quantum correlations in the thermometric performance, which emerge as a central feature of this work. 
We find strong numerical evidence that, given same single-qubit reduced states, the inclusion of quantum correlations among the qubits of the ensemble always yields an enhanced QFI. Moreover, even if none of the considered states outperform the (pure, separable) ground state, maximally entangled states display QFIs values remarkably close to the standard quantum limit when probing extremely hot thermal baths.
Finally, although the Markovian dynamics does not permit superlinear scaling of the QFI with the number of probes, we identify the most effective initial states for designing high-precision quantum thermometers within this setting. We also provide concrete guidelines for experimental implementations. 
\end{abstract}

\maketitle

\section{Introduction}

Quantum metrology can offer significant advantages over its classical counterpart~\cite{paris2009quantum,giovannettifirst,giovannetti2011advances}, as it allows for an extremely high spatial resolution and an improved precision on parameter estimations. In terms of the former, one can engineer quantum sensors at the atomic scale to allow high-spatial resolution~\cite{DegenRMP2017}. Moreover, metrological performance can be directly enhanced by exploiting genuine quantum resources, most notably the fragile yet highly sensitive character of quantum superposition and quantum entanglement. For example, quantum gravitational wave detectors~\cite{Aasi_2013_short} use squeezed states of light to enhance their sensitivity to detect extremely weak signals. Furthermore, when operating in a multi-probe setting, quantum metrology allows for a superlinear scaling of the precision with the number of probes~\cite{montenegro2024reviewquantummetrologysensing}.

\begin{figure}[t]
    \centering
    \includegraphics[width=0.9\linewidth]{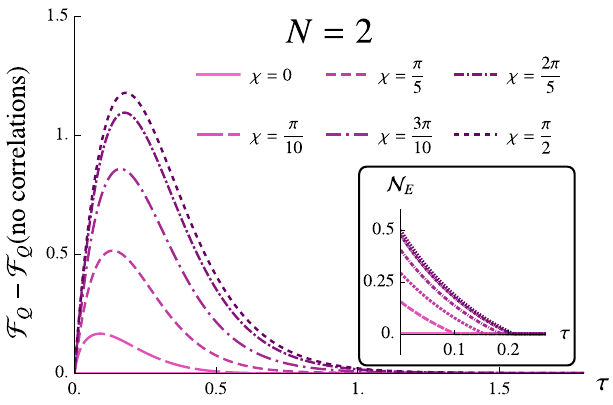}
    \caption{
    Enhancement of estimation precision due to quantum correlations of an input two-qubit ($N=2$) probe state. We compare the QFIs of the probe state undergoing thermalization over time $\tau$ (taken here dimensionless), initializing the probe in a squeezed spin state and in the product state of its reduced states, respectively. The parameter $\chi$ regulates the amount of correlations, which is quantified by the entanglement negativity $\mathcal{N}_E$, shown in the inset. }
    \label{fig:sq_minus_kronprod}
\end{figure}

Analogous advantages emerge also in quantum thermometry, where quantum probes are employed as thermometers to estimate the temperature of a macroscopic thermal bath with high precision~\cite{MahboudiJPA2019,Abiuso_2024}. In this setting, the key physical quantity used to characterize the precision of temperature estimation at any time $t$ is the Quantum Fisher Information (QFI) associated with the state of the thermometers. This is exactly the quantity displayed in Fig.~\ref{fig:sq_minus_kronprod} for a relevant case-study in our analysis.  Within the quantum thermometry context, the information about the temperature of the thermal bath is encoded in the dissipative part of the thermometers' dynamics responsible for their thermalization. Among the first models of thermometry in the quantum regime we mention~\cite{brunelli_2011,BrunelliPRA2012} where a qubit is used to probe the temperature of a micromechanical resonator, modelled as a quantum harmonic oscillator. This scenario has been also further explored in the nonlinear optomechanical regime~\cite{Montenegro_2020}. Another successful approach to quantum thermometry is via collisional models, in which a small quantum probe estimates the temperature of a target environment through repeated, discrete interactions rather than continuous coupling~\cite{Seah_2019, e23121634,Mendon_a_2025,alves_landi_2024}. In this paper, we set our analysis in the context of qubit thermometers~\cite{JevticPRA2015, RazavianEPJP2019}, which have been experimentally tested in Refs.~\cite{ThamSciRep2016,MancinoPRL2017} on a quantum optics platform. Other notable examples include non-destructive thermometry of a Bose-Einstein condensate using impurities as quantum thermometer~\cite{Sab_n_2014}, thermometry of ultracold atoms via nonequilibrium work distributions~\cite{Johnson_2016}, non-destructive thermometry of cold Fermi gases~\cite{Mitchison_2020}, thermometry of superconducting quantum circuits~\cite{Sultanov_2021}, pairs of trapped ions used as thermometers~\cite{de_S_Neto_2022}, and even biological applications of quantum thermometry with cells~\cite{yang2011quantum,wu2022recent}. 

Recent studies~\cite{PhysRevE.110.024132,Vasco2018,Fraz_o_2024, alves_landi_2022, marti_paper} have highlighted the advantages of measurement strategies conducted in nonequilibrium regimes, where the quantum probes are measured before reaching their (local) thermal equilibrium. This approach yields a twofold benefit. First, the procedure is faster than equilibrium-based thermometry, since the probes are measured before full thermalization occurs. Second, it can lead to markedly improved precision in estimating the bath’s temperature, owing to a QFI peak higher than its asymptotic value.

Quantum resources play a central role in enabling metrological advantages within the considered framework. For example, in Ref.~\cite{Fraz_o_2024} the authors demonstrated that the presence of initial quantum coherence can boost the value of the QFI, thus leading to improved temperature sensitivity at finite times. Moreover, in Ref.~\cite{Seah_2019}, a collisional thermometry framework is introduced, where a quantum probe interacts sequentially with a series of $N$ auxiliary systems that correlate over time. The information about the temperature of the bath is extracted from the state of the auxiliary systems whose QFI is shown to scale superlinearly with $N$. Furthermore, Ref.~\cite{Ravell_Rodr_guez_2024} investigates quantum thermometry using fermionic probes strongly coupled to a thermal bath. They show how the induced non-Markovian dynamics enhances the achievable QFI compared to its Markovian counterpart, possibly leading to a superadditive QFI for a two-fermion quantum probe. Ref.~\cite{Salvatori_2014} discusses a quantum thermometry protocol by means of a thermalizing Lipkin-Meshkov-Glick critical probe of up to four particles. By tuning the anisotropy and the external magnetic field, they demonstrate that it is possible to saturate the ultimate bound on the precision of thermometry.

Despite recent advances, the impact of multipartite correlations in the state of the thermometers on nonequilibrium quantum thermometry is not yet fully understood and remains an open research question~\cite{Mirkhalaf_2024}. Interestingly, the presence of entanglement does not necessarily imply an operational advantage: for instance, when the goal is to estimate the parameter of an amplitude-damping channel (only spontaneous emission is modeled), the presence of entanglement can be detrimental~\cite{fujiwara2004estimation}. Moreover, in general, there is no guarantee that a single input state performs optimally for all values of the parameter to be estimated~\cite{MonrasPRL2007,shu_2020}. For the depolarizing channel~\cite{fujiwara2001quantum,boixo2008operational} the optimal state is the maximally entangled one for small values of the depolarizing parameter, whereas for values larger that a given threshold the optimal choice becomes a separable (pure) state (see Fig.~1 of Ref.~\cite{boixo2008operational}). Even the optimal measurement basis for metrological purposes may depend on the value of the parameter and/or be highly non-local~\cite{RubioPRL2021,prl_Mehboudi,prx_Glatthard}. We emphasize that it crucial to determine the operational regimes in which quantum-correlated probes offer a genuine advantage: on the one hand, preparing correlated states is experimentally resource-intensive; on the other hand, in certain scenarios such correlations constitute the most valuable resource.

In this paper we aim to understand the conditions under which the precision of nonequilibrium bath temperature estimation is enhanced by employing an ensemble of qubits prepared in a quantum-correlated state. To do this, we consider the Generalized Amplitude Damping (GAD) quantum channel~\cite{nielsen2010quantum}, which contains both absorption and spontaneous emission. Then, we introduce relevant correlated initial states for the ensemble of thermometers (described in detail in Sec.~\ref{sec:input_states} and listed in Tab.~\ref{tab:states_summary}), whose features are representative of the mechanisms that lead to the best achievable precision. In our analysis, we focus on nonequilibrium thermometry conducted during the transient of the qubits' dissipative dynamics induced by the weak coupling with the thermal bath.

For uncorrelated GAD channels, we provide strong numerical evidences that the presence of correlations among individual probes in their initial state before the interaction with the bath is always associated with an enhancement of the QFI, if we compare with the tensor product of the single-probe reduced states. This becomes evident in Fig.~\ref{fig:sq_minus_kronprod}, where we display the difference between the QFIs of both the squeezed spin state and the tensor product of its reduced states. The difference in the transient QFIs in the plot is entirely ascribable to quantum correlations.

Furthermore, we find that the thermometry is more precise when the thermal bath is relatively hot, in agreement with the existing literature~\cite{shu_2020,mukherjee2019enhanced,correa_2017,zhang2024enhancinglowtemperaturequantumthermometry,Potts2019fundamentallimits,ullah_2023}. Within the range of parameters explored, also a low local temperature of the individual thermometers is beneficial for enhancing precision. Then, once the local populations of each thermometer are fixed, the presence of local coherence consistently improves the estimation accuracy.
Concerning the presence of correlations, when the temperature of the bath is sufficiently high, the maximally correlated state achieves a time-optimized QFI that is remarkably close to that of the best performing separable input states. Our numerical results, indeed, show that---among the states we analyze---the best-performing ones are separable. In particular, the highest transient peak of the QFI is achieved by the ground state of the reference Hamiltonian, in agreement with the existing literature~\cite{shu_2020,marti_paper}. However, at later times---though still before full thermalization---the optimal state shifts to a product state endowed with substantial local coherence. The described behaviors demonstrates that, for the quantum thermometry protocol considered in this paper, quantum correlations can represent a valuable resource to a higher precision in estimating the bath temperature.

We also analyze the behavior of the QFI as a function of the number $N$ of thermometers. By leveraging the ultimate bound~\cite{Fujiwara_2008,Demkowicz_Dobrza_2013} on the QFI for the estimation of the bath temperature that is encoded in uncorrelated GAD channels, we prove that a superlinear asymptotic scaling with $N$ cannot be achieved. Nevertheless, we examine how the time-optimized QFI scales with $N$ for various choices of the input state. We also investigate the behavior of the corresponding optimal time at which an interrogation of the probes yields the maximal information on the bath temperature. Consistent with earlier results, the most favorable scaling is achieved by preparing the thermometers in the ground state. However, we show that correlations bring an enhancement on the scaling with $N$ of the QFI, again when comparing the results with those obtained by initializing the thermometers in the tensor product of the single-thermometer reduced states.

We believe that our results, albeit much of them numerical and valid for specific values of $\beta, N$, can provide a useful framework for an experimental implementation of the considered thermometry protocols, giving concrete guidelines on the optimal input states and measurement times.

\section{Model}

\begin{figure}[t]
    \centering
    \includegraphics[width=0.55\linewidth]{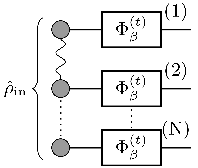}
    \caption{\justifying Quantum-map representation of the system-environment interaction. The system consists of an ensemble of $N$ qubits operating as quantum thermometers. The environment is a thermal bath with unknown inverse temperature $\beta$. Each qubit thermometer undergoes thermalization governed by the local map $\Phi_\beta^{[t]}$. The ensemble is initially prepared in a quantum-correlated state $\hat{\rho}_{\rm in}$.}
    \label{fig:system_environment_setup}
\end{figure}

The physical setting considered in this paper is pictorially represented in Fig.~\ref{fig:system_environment_setup}. The probe system is composed of $N$ qubits and each of them is a quantum thermometer that is used for estimating the temperature of a thermal bath. The latter is characterized by the inverse temperature $\beta \equiv 1/(k_B T)$, where the Boltzmann constant $k_B$ is fixed to $1$ from here on for convenience.

We allow for the $N$ qubits to be initially prepared in a correlated state, but we assume that no interactions occur among them during the thermalization. Therefore, each qubit evolves independently and interacts only with the thermal bath during the dynamics.

We now explain how the time evolution of each individual qubit is modeled. Each qubit thermalizes independently in the basis formed by the eigenvectors of the following Hamiltonian
\begin{equation}\label{eq:hamiltonian}
    \hat{H} = -\frac{\hbar\omega}{2}\hat{\sigma}_z.
\end{equation}
The ground state $|0\rangle$ has energy $\varepsilon_0 = -\hbar\omega/2$ and the excited state $|1\rangle$ energy $\varepsilon_1 = \hbar\omega/2$. A thermal state at inverse temperature $\mu$ is represented, in the Hamiltonian eigenbasis, by the diagonal matrix 
\begin{equation}
    \thermal(\mu) =
    \begin{pmatrix}
        \pi_0(\mu) & 0 \\
        0 & \pi_1(\mu)
    \end{pmatrix},
    \label{eq:thermal_state}
\end{equation}
where $\pi_0(\mu)=e^{-\mu\epsilon_0}/Z_{\mu}$ is the thermal population of the ground state and $\pi_1(\mu)=e^{-\mu\epsilon_1}/Z_{\mu}$ that of the excited state. Here, $Z_\mu = e^{-\mu \varepsilon_0}+e^{-\mu \varepsilon_1}$ is the qubit's partition function. 

We assume that each qubit interacts weakly with the thermal bath and, because of that, its state evolves via a genuinely dissipative Markovian quantum map $\Phi_\beta^{[t]}$. Under the action of this map, the thermalization decay rate is
\begin{equation}
    \lambda = \frac{\gamma}{\pi_0(\beta) - \pi_1(\beta)},
    \label{eq:lambda}
\end{equation}
which is positive if $\beta>0$. In Eq.~\eqref{eq:lambda}, the parameter $\gamma$ quantifies the coupling strength between the qubit and the bath, and thus determines the rate at which thermalization occurs. Further details are provided in Appendix~\ref{sec:appendix_master_equation}.

The quantum map $\Phi_\beta^{[t]}$ can be effectively represented through a GAD quantum channel, realized by the set of Kraus operators $\mathcal{K} \equiv \{ \hat{K}_1, \hat{K}_2, \hat{K}_3, \hat{K}_4 \}$, where
\begin{eqnarray}
    \hat{K}_1 &=& \sqrt{q}
        \begin{pmatrix}
            1 & 0 \\ 0 & \sqrt{1-p}
        \end{pmatrix},
        \qquad \quad
        \hat{K}_2 = \sqrt{q}
        \begin{pmatrix}
            0 & \sqrt{p} \\ 0& 0
        \end{pmatrix}, 
        \label{eq:GAD_Kraus_operators}
        \\
        \hat{K}_3 &=&\sqrt{1-q}
        \begin{pmatrix}
            \sqrt{1-p} & 0 \\ 0 & 1
        \end{pmatrix},
        \quad
        \hat{K}_4  =\sqrt{1-q}
        \begin{pmatrix}
            0 & 0 \\ \sqrt{p} & 0 
        \end{pmatrix}   \nonumber
\end{eqnarray}
Here, $q = \pi_0(\beta)$ and $p=1-e^{-\lambda t}$ are dynamical parameters that depend on the inverse temperature $\beta$ of the bath. Under the action of this channel, an input single-qubit state $\hat{\rho}_{\rm in}^{(1)}$ evolves to $\Phi_\beta^{[t]}[\hat{\rho}_{\rm in}^{(1)}]= \sum_a \hat{K}_a \hat{\rho}_{\rm in}^{(1)} \hat{K}_a^\dagger$, providing the output state 
\begin{equation}
\hat{\rho}_{\rm out}^{(1)}(t,\beta) 
= \begin{pmatrix}
        e^{-\lambda t} (\rho_{11}- \pi_0 ) & e^{-\frac{\lambda}{2}t} \rho_{12}  \\
        e^{-\frac{\lambda}{2}t} \rho_{21} &   e^{-\lambda t} (\rho_{22} - \pi_1)
    \end{pmatrix}
    + \hat{\rho}_{\rm th}(\beta),
\label{eq:gad_kraus}
\end{equation}
where $\rho_{jk}$ are the matrix elements of $\hat{\rho}_{\rm in}^{(1)}$. As shown by Eq.~\eqref{eq:gad_kraus}, all input states $\hat{\rho}_{\rm in}^{(1)}$ evolve asymptotically towards the thermal state $\hat{\rho}_{\rm th}(\beta)={\rm diag}[\pi_0(\beta),\pi_1(\beta)]$ at bath temperature, namely the fixed point of the GAD channel. We thus have
\begin{equation}
    \lim_{t \to + \infty} \Phi_\beta^{[t]}[\hat{\rho}_{\rm in}^{(1)}] = \hat{\rho}_{\text{th}} (\beta).
    \label{eq:single_qubit_lim}
\end{equation} 

Having characterized the single-qubit dynamics, we now consider an ensemble of $N$ qubits acting as quantum thermometers. Since each qubit evolves independently from the others, the overall dynamical map is obtained by applying $N$ uncorrelated GAD quantum channels:
\begin{equation}
    \hat{\rho}_{\text{out}}(t,\beta) = \left(\Phi_\beta^{[t]}\right)^{\otimes N}[\hat{\rho}_{\text{in}}],
    \label{eq:rho_out}
\end{equation}
where $\hat{\rho}_{\text{in}}$ and $\hat{\rho}_{\text{out}}(t,\beta)$ are the input and evolved states of the ensemble, respectively. By combining Eqs.~\eqref{eq:single_qubit_lim} and \eqref{eq:rho_out}, and taking into account the Markovianity of the dynamics, we find
\begin{equation}\label{eq:thermalization}
    \lim_{t \to + \infty} \left( \Phi_\beta^{[t]} \right) ^{\otimes N } [\hat{\rho}_{\text{in}}] = \hat{\rho}_{\text{th}}^{\otimes N} (\beta).
\end{equation}
This entails that every initial state $\hat{\rho}_{\text{in}}$ asymptotically tends to the tensor product of $N$ single-qubit thermal states at the inverse temperature of the bath. Due to the independence of the quantum channels applied to the qubits, no correlation can emerge during the time evolution. Moreover, as the quantum map of Eq.~\eqref{eq:gad_kraus} is Markovian, quantum correlations in the initial state of the ensemble vanish exponentially over time.

\section{Quantum Fisher Information}
\label{sec:QFI}

The Quantum Fisher Information (QFI) is a central tool to assess the maximum achievable precision in the estimation of a parameter---the inverse bath temperature $\beta$ in our case---whose unknown value is encoded into the state of the quantum system used as a probe~\cite{Helstrom1967,Liu2019}. Specifically, the QFI enters the Quantum Cram\'{e}r-Rao bound (QCRB)~\cite{Helstrom1967,pluto}
\begin{equation}
    \Delta \beta \ge \frac{1}{\sqrt{\nu \mathcal{F}_Q\big(\hat{\rho}_{\text{out}}(t,\beta)\big)}},
    \label{eq:uncertanty_on_beta_estimation}
\end{equation}
where $\mathcal{F}_Q\big(\hat{\rho}_{\text{out}}(t,\beta)\big)$ is the QFI of the evolved state in Eq.~\eqref{eq:rho_out}, and $\nu$ is the number of independent measurement repetitions. Eq.~\eqref{eq:uncertanty_on_beta_estimation} sets a fundamental limit on the standard deviation $\Delta \beta$ achievable in the estimation of $\beta$.

The QFI is defined through the Symmetric Logarithmic Derivative (SLD), denoted by $\hat{L}_\beta (t)$, which is implicitly given by the Lyapunov equation  
\begin{equation}
     \frac{\partial \hat{\rho}_{\text{out}}(t,\beta)}{\partial \beta} = \frac{1}{2} \big\{ \hat{\rho}_{\text{out}}(t,\beta), \hat{L}_\beta(t) \big\}.
     \label{eq:lyapunov}
 \end{equation}
In terms of the SLD, the QFI can be expressed as
\begin{equation}
    \mathcal{F}_Q \big( \hat{\rho}_{\text{out}}(t,\beta) \big) = \tr \left[ \frac{\partial \hat{\rho}_{\text{out}}(t,\beta)}{\partial \beta} \hat{L}_\beta(t) \right]. 
\label{eq:qfi_def} 
\end{equation} 

As indicated by Eq.~\eqref{eq:uncertanty_on_beta_estimation}, it is desirable to maximize $\mathcal{F}_Q\big(\hat{\rho}_{\text{out}}(t,\beta)\big)$ in order to minimize the estimation uncertainty. Since the parameter-encoding map $\Phi_\beta^{[t]}$ depends on $t$, different stages of the thermalization process may correspond to a more or less favorable sensitivity. For $t=0$, we have 
\begin{equation}
    \mathcal{F}_Q \big(\hat{\rho}_{\text{out}}(t=0,\beta)\big)=0,
\end{equation}
as the map $\Phi_\beta^{(t=0)}=\id$ provides no information about $\beta$. On the other hand, for $t\to +\infty$, all input states are mapped to a product of thermal states at inverse bath temperature $\beta$, according to Eq.~\eqref{eq:thermalization}. Using the additivity of the QFI on product states and the relation $\mathcal{F}_Q (\hat{\rho}_{\text{th}}(\beta))=\Delta^2\hat{H}$~\cite{Vasco2018,Fraz_o_2024,CitekeyMisc1}, where $\Delta^2\hat{H}$ is the thermal variance at inverse temperature $\beta$ of the Hamiltonian in Eq.~\eqref{eq:hamiltonian} (see Appendix~\ref{sec:qfi_thermal} for details), we find 
\begin{align}\label{eq:QFI_thermal}
    \lim_{t\to+\infty}\mathcal{F}_Q \big(\hat{\rho}_{\text{out}}(t,\beta)\big)=N(\varepsilon_1 - \varepsilon_0)^2 \pi_0(\beta)\,\pi_1 (\beta).
\end{align}
Therefore, at thermal equilibrium, the ultimate achievable precision given by the QCRB is limited by the bath temperature through the thermal populations $\pi_0(\beta)$ and $\pi_1(\beta)$. Initial quantum correlations between qubits are of no advantage in this regime, as they are completely washed out by the thermalization process. This motivates our investigation of the non-equilibrium transient regime, where larger QFI than Eq.~\eqref{eq:QFI_thermal} can in fact be attained.

\subsection*{GAD sensitivity limits}
\label{sec:qfi_bound}

The thermometer initial state, $\hat{\rho}_{\rm in}$, also plays a major role in determining the QFI behavior. A relevant class of such states will be introduced in Sec.~\ref{sec:input_states}, and their impact on sensitivity will be examined in Sec.~\ref{sec:results}. Here, however, we are only concerned with sensitivity limits that do not depend on the specific choice of $\hat{\rho}_{\rm in}$. Rather, these arise as intrinsic limits of the encoding channel. To better illustrate this point, we define the \textit{channel QFI}~\cite{Demkowicz_Dobrza_ski_2012}
\begin{equation}
\mathcal{F}_Q\left(\Phi_\beta^{[t]}\right) = \max_{\hat{\rho}_{\text{in}}} \mathcal{F}_Q \left( \left(\Phi_\beta^{[t]}\right)^{\otimes N}\left[ \hat{\rho}_{\text{in}} \right] \right),
\label{channel_qfi} 
\end{equation}
which, in our case, represents the maximum QFI achievable with $N$ independent GAD encoding channels. Instead, if we do not allow quantum correlations in the input state, the maximum is always achieved by a product state $(\hat{\rho}_{\rm in}^{(1)})^{\otimes N}$ constructed from the state $\hat{\rho}_{\rm in}^{(1)}$ that maximizes the single-qubit QFI~\cite{topolino}. We thus obtain the Standard Quantum Limit (SQL) of the channel
\begin{equation}
    \mathcal{F}_Q^{\rm SQL}\left(\Phi_\beta^{[t]}\right)=N\max_{\hat{\rho}_{\rm in}^{(1)}}\mathcal{F}_Q \left(\Phi_\beta^{[t]}\left[ \hat{\rho}_{\text{in}}^{(1)}\right]\right),
    \label{SQL}
\end{equation}
whose scaling is always linear with $N$. 

For unitary encoding channels, the SQL can be surpassed by the channel QFI in Eq.~\eqref{channel_qfi}, which yields a quadratic scaling with $N$, the \textit{Heisenberg scaling}~\cite{giovannetti2001quantum}. Here, a superior sensitivity is achieved due to highly entangled states similar to the GHZ state of Eq.~\eqref{GHZ} (see below). This analysis breaks down when the encoding channel is non-unitary. This corresponds to our case study, as the inverse temperature $\beta$ of the thermal bath is encoded in the Kraus operators of Eq.~\eqref{eq:GAD_Kraus_operators}. In some cases, a gain over the SQL is still possible, but this reduces to a constant factor with respect to $N$~\cite{Demkowicz_Dobrza_ski_2012}. In other cases, entanglement may provide no advantage. Interestingly, Fujiwara has highlighted the detrimental role played by entanglement in a channel-estimation problem closely related to GAD~\cite{Fujiwara2004}.

The optimization in Eq.~\eqref{SQL} can be solved based on the analytic expression of the QFI for the single qubit~\cite{Fraz_o_2024}. Due to the convexity of the QFI, the maximum is achieved for some pure input state $|\psi_{\rm in}^{(1)}\rangle=\sqrt{1-a}|0\rangle+\sqrt{a}e^{i\phi}|1\rangle$. 
\begin{figure}[t]
    \centering
    \subfloat[]{%
    \includegraphics[width=0.475\textwidth]{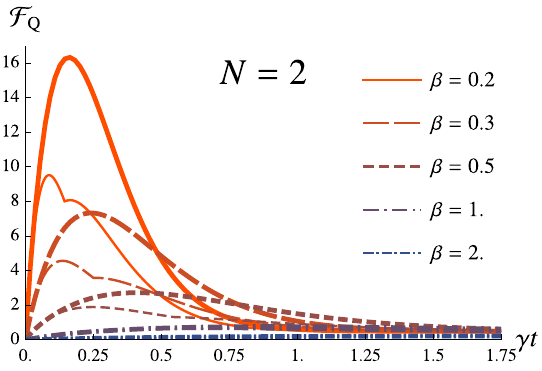}
        \label{fig:first}
    }
    \vspace{-2mm}
    \subfloat[]{\includegraphics[width=0.475
    \textwidth]{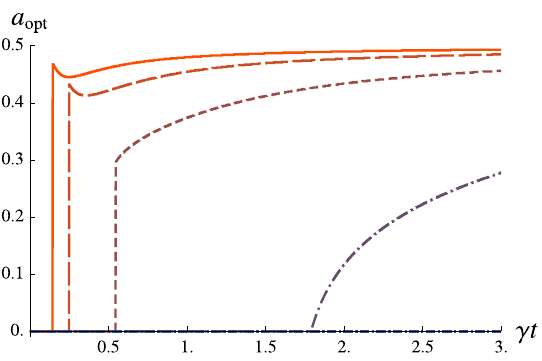}
        \label{fig:second}
    }
    \caption{
    \justifying{Panel (a): SQL, Eq.~\eqref{SQL} (thin curves), and upper bound, Eq.~\eqref{eq:bound} (thick curves). Panel (b): optimal values $a_{\rm opt}$ for the parameter $a$ that defines the single-qubit input state $|\psi_{\rm in}^{(1)}\rangle=\sqrt{1-a}|0\rangle+\sqrt{a}|1\rangle$. The values of $a_{\rm opt}$ identify those input states that provide the maximum QFI achievable with no quantum correlations at different times.}
    }
    \label{fig:bound_figure}
\end{figure}
Here, since the QFI is independent of $\phi$, $0\leq a\leq 1$ is the only free parameter to optimize.

The optimization in Eq.~\eqref{channel_qfi} is more challenging to perform as the parameter space grows exponentially with the number of qubits. In Appendix \ref{sec:m2_is_null}, we prove the validity of the following upper bound~\cite{Demkowicz_Dobrza_ski_2012}:
\begin{equation}
\mathcal{F}_Q\left(\Phi_\beta^{[t]}\right) \leq 4 N \min_{\hat{h}}  \norm{\hat{A}_{\mathcal{K}'}} ,
    \label{eq:bound}
\end{equation}
where the symbol $\norm{\cdot}$ is the operator norm. Here, $\mathcal{K}'$ denotes an equivalent set of Kraus operators, obtained from Eq.~\eqref{eq:GAD_Kraus_operators} by means of $\hat{K}'_j=\sum_{l=1}^4(e^{-i\beta \hat{h}})_{jl}\hat{K}_l$ ($j=1,\dots,4$), and $A_{\mathcal{K}'}=\sum_{l=1}^4(\partial\hat{K}'_l/\partial\beta)^{\dagger}(\partial\hat{K}'_l/\partial\beta)$. In the above equations, $\hat{h}$ denotes the generic element of a set of $4\times4$ Hermitian matrices that satisfy a given linear constraint (see Appendix~\ref{sec:m2_is_null} for details). From a computational point of view, the optimization on $\hat{h}$ can be performed efficiently as the dimension of the parameter space is independent of $N$. Eq.~\eqref{eq:bound} shows that the asymptotic scaling of the channel QFI must be linear with $N$, thereby ruling out the possibility of achieving Heisenberg scaling. The limitation lies in the encoding via uncorrelated GAD channels, not in any specific input state.

Results obtained by performing the optimizations in Eqs.~\eqref{SQL} and~\eqref{eq:bound} are shown in Fig.~\ref{fig:bound_figure} panel (a), where thin curves represent the SQL and thick curves the upper bound. For small values of $\beta$ (i.e., high bath temperatures), two closely spaced local maxima characterize the SQL during the transient regime, both surpassing the QFI asymptotic value given by Eq.~\eqref{eq:QFI_thermal}. The first and highest maximum is obtained for $a=0$, corresponding to the product state $|0\rangle^{\otimes N}$, which is the optimal input state at early times in the absence of quantum correlations. For later times, however, the optimal input state is a function of the time elapsed since the beginning of thermalization. The optimal $a$ value falls inside the interval $0<a<1/2$, corresponding to a product state with nontrivial local coherence (see Sec.~\ref{sec:input_states} below), which generates the second, lower maximum. The abrupt transition between the two types of optimal states is marked by a discontinuity in the SQL derivative in panel (a) and by a jump in the optimal value of $a$ in panel (b). For large values of $\beta$ (i.e., low bath temperatures), no peak emerges; instead, the largest value of the QFI is attained asymptotically. An analogous behavior is observed for the upper bound (thick curves), except that only a single peak of the QFI appears during the transient regime.

When comparing the SQL with the upper bound, we find that the smaller the value of $\beta$, the larger the discrepancy between the two bounds. This opens up the possibility of using entangled states to surpass the SQL, especially for low values of $\beta$ and at early times. We investigate this possibility in the following sections.

 \begin{table*}[t]
    \centering
    \begin{tabular}{p{1.7cm}|p{1.3cm}|p{4.3cm}|p{3.2cm}|p{6.5cm}}
        \textbf{Symbol} & \textbf{Def.} & \textbf{Name} & \textbf{Parameters} & \textbf{What the parameters tune (resources)} \\
        \hline 
        \hline
        $\ket{0}^{\otimes N}$ & - & Ground state & - & - \\
         \hline
         $\ket{{\rm GHZ}}$ & Eq. \eqref{GHZ} & GHZ state & - & - \\
        \hline
        $\thmix(\eta,\mu)$ & Eq. \eqref{eq:thermal_ghz_mix} & Thermal-mixtured GHZ state modified by thermal weights & $\eta \in [0,1];\, \mu \in [0,+\infty[$ & $\eta$: quantum correlation, $\mu$: local inverse temperature, quantum correlation \\
        \hline 
        $\idmix (\eta)$ & Eq. \eqref{eq:ghz_mix} & Identity-mixtured GHZ state & $\eta \in [0,1]$ & $\eta$: quantum correlation \\
        \hline
        $\ket{\psi_{\text{k-s}} (\alpha) }$ & Eq. \eqref{eq:k_GHZ_superposition} & $k$-GHZ superposition state & $\alpha \in [0, \pi[$ & $\alpha$: quantum correlations (local coherence when $\ket{k} = \ket{\pm}$) \\
        \hline
        $\ket{\psi_{\rm sq} (\chi)}$ & Eq. \eqref{eq:squeezed_state} & Squeezed spin state & $\chi \in [0,\pi[$ & $\chi$: quantum correlation, local coherence
    \end{tabular}
    \caption{Summary of the input states $\hat{\rho}_{\rm in}$ employed in this paper.}
    \label{tab:states_summary}
\end{table*}

\section{Correlated input states} 
\label{sec:input_states}

In this work, we consider a class of $N$-qubit input states $\hat{\rho}_{\text{in}}$ that exhibit relevant quantum properties, such as quantum coherence or entanglement. Each state is characterized by specific parameters that determine the extent to which these properties are present in the state. This enables a systematic analysis of how different quantum resources influence the achievable precision in the estimation of $\beta$.

All the states considered in this section are invariant under particle exchange. Consequently, for each density operator $\hat{\rho}$ listed below, the compact notation $\tr_{N-1}[\hat{\rho}]$ refers to the single-qubit reduced state obtained by tracing over any arbitrary subset of $N-1$ qubits. We can also define the \textit{local inverse temperature} and the \textit{local coherence} of a generic $N$-qubit state in terms of the matrix elements of its single-qubit reduced density operator. The diagonal elements, $\rho_{22}$ and $\rho_{11} = 1 - \rho_{22}$, determine the local inverse temperature as  
\begin{equation}
\mu = \frac{1}{\varepsilon_1 - \varepsilon_0} \ln\left( \frac{\rho_{11}}{\rho_{22}} \right), 
\label{eq:local_inverse_temperature_mu}
\end{equation}
by analogy with the thermal case, while the module $|\rho_{12}|=|{\rho_{21}}|$ of the off-diagonal elements quantifies the local coherence of the state.

A summary of the states introduced below can be found in Table \ref{tab:states_summary}.

\paragraph{Thermal-mixtured GHZ states modified by thermal weights ($\text{th-m}$).} 
Let us construct the superposition state
\begin{equation}
    \ket{\psi_{\text{th-m}} (\mu) }  \equiv \sqrt{\pi_0(\mu)}  \ket{0}^{\otimes N} + \sqrt{\pi_1 (\mu)}  \ket{1}^{\otimes N},
    \label{eq:th_superposition}
\end{equation}
with probabilities given by the populations $\pi_0(\mu)$ and $\pi_1(\mu)$ of the thermal state $\hat{\rho}_{\rm th}(\mu)$. Then, we define the target density operator as a mixture of $\ket{\psi_{\text{th-m}}(\mu)}$ and the tensor product of $N$ thermal states at inverse temperature $\mu$:
\begin{equation}\label{eq:thermal_ghz_mix}
    \thmix (\eta, \mu) = \eta \ketbra{\psi_{\text{th-m}}(\mu)}{\psi_{\text{th-m}}(\mu)} + (1-\eta) \hat{\rho}_{\text{th}}^{\otimes N}(\mu),    
\end{equation}
where $\eta \in [0,1]$. The parameters $\eta$ and $\mu$ tune the amount of correlations among qubits, which grows as $\eta$ increases and is reduced when $\mu$ decreases. The corresponding single-qubit reduced state remains unaffected by the value of $\eta$:
\begin{equation}
          \tr_{N-1} \left[ \thmix (\eta, \mu) \right] = \hat{\rho}_{\text{th}}(\mu), 
\end{equation}
whereby the parameter $\mu$ corresponds to the local inverse temperature of the individual qubits according to Eq.~\eqref{eq:local_inverse_temperature_mu}.

When $\mu \to +\infty$, the ground state $\ket{0}^{\otimes N}$ is recovered. On the other hand, when $\mu \to 0^+$, the superposition in Eq.~\eqref{eq:th_superposition} becomes the well-known GHZ state~\cite{greenberger2007goingbellstheorem,greenberger1990bell}
\begin{equation}
   \ket{\text{GHZ}} = \frac{\ket{0}^{\otimes N} + \ket{1}^{\otimes N}}{\sqrt{2}}.
   \label{GHZ}
\end{equation}
Thus, the density operator in Eq.~\eqref{eq:thermal_ghz_mix} simplifies to the state
\begin{equation}
    \idmix (\eta) = \eta \ket{\text{GHZ}} \bra{\text{GHZ}} + (1-\eta)\frac{\mathbb{1}_{2^N}}{2^N}, 
    \label{eq:ghz_mix}
\end{equation}
which we call the \textit{identity-mixtured GHZ state}. Interestingly, when $N=2$, $\idmix$ corresponds to an isotropic state~\cite{horodecki1998reductioncriterionseparabilitylimits}, which in turn is related to a Werner state~\cite{werner_states} since applying the partial transposition operation to one of the two subsystems (qubits in our case) transforms an isotropic state to a Werner state and vice versa~\cite{osti_23183265}.

\paragraph{$k$-GHZ superposition states (k-s).} These states are constructed as a superposition of the $\ket{\text{GHZ}}$ state and the product state $\ket{k}^{\otimes N}$, where $\ket{k}$ is a generic single-qubit pure state. The weights of the superposition are given by $\sin{(\alpha)}$ and $\cos{(\alpha)}$, respectively, divided by the appropriate normalization constant $C(k,N,\alpha)$ that depends on $\alpha$, $k$ and the number $N$ of qubits. Hence,
\begin{equation}
        \ket{\psi_{\text{k-s}} (\alpha) } = \frac{\sin (\alpha) \ket{\text{GHZ}} + \cos(\alpha) \ket{k}^{\otimes N}}{C( k,N,\alpha)};
        \label{eq:k_GHZ_superposition}
\end{equation}
the density operator associated to this pure state is $ \kghz (\alpha) = \ket{\psi _{\text{k-s}}(\alpha) } \bra{\psi_{\text{k-s}} (\alpha) }$.
The parameter $\alpha$ determines the correlations among the qubits—being maximal for the GHZ state and vanishing when the state is fully separable—but it also affects other properties as local inverse temperature and coherence. For instance, by setting $\ket{k} = \ket{+} = (\ket{0} +\ket{1})/\sqrt{2}$ (as we will do below), $\alpha$ controls the local coherences of the input state, while the corresponding populations are constant and equal to $1/2$. In this case, the reduced density operators is indeed
\begin{equation}
    \tr_{N-1} [\plusghz (\alpha)] = \frac{1}{2}
    \begin{pmatrix}
        1 & \hspace{-5mm} 1 - x(\alpha,N) \\
       1 - x(N,\alpha) &  1
    \end{pmatrix},
    \label{eq:reduced_plusghz_state}
\end{equation}
where $x(N,\alpha) = \sin^2(\alpha)/ [1 + 2^{\frac{1-N}{2}}\sin(2\alpha)]$; the details of the derivation of Eq.~\eqref{eq:reduced_plusghz_state} are given in Appendix~\ref{sec:reduced_plusghz}.

\paragraph{Squeezed spin states.} ``Spin squeezing'' refers to the reduction of quantum fluctuations in a collective pseudo-spin observable within a finite-dimensional system. Interestingly for us, this effect is associated with the generation of entanglement among the system's constituents. 

Squeezed spin states have been realized through different techniques on a variety of experimental platforms, and play a central role in quantum metrology~\cite{Pezze2018,Ma2011,Eckner2023}. In the following, we consider squeezed spin states generated via one-axis twisting~\cite{Kitagawa1993}. If the $N$ qubit thermometers are initially prepared in the product state $\ket{+}^{\otimes N}$, one-axis twisting
\begin{equation}
    \squeezvec{\chi} = e^{- i \chi \hat{J}_z^2} \ket{+}^{\otimes N}
    \label{eq:squeezed_state}
\end{equation}
generates a squeezed spin state with spin squeezing in the $\hat{J}_x-\hat{J}_y$ plane. Here, the triplet $\hat{J}_{\alpha} \equiv \sum_{i=1}^N \sigma^{(i)}_{\alpha}/2$ ($\alpha=x,y,z)$ satisfies the $\mathfrak{su}(2)$ commutation relations, and can therefore be interpreted as a set of pseudo-spin operators. The squeezing strength is determined by the parameter $\chi$. The state is periodic in $\chi$ up to a phase factor: we have a complete revival of the initial state for $\chi=\pi$ if $N$ is odd, and for $\chi=2\pi$ if $N$ is even. 

Using $\squeez (\chi)  \equiv \ketbra{\psi_{\rm sq} (\chi)}{\psi_{\rm sq} (\chi)}$, the reduced density operator for each qubit is found to be
\begin{equation}
    \tr_{N-1} \left[ \squeez (\chi) \right] 
     =\frac{1}{2}
     \begin{pmatrix}
         1 & \hspace{-3mm} \cos^{N-1}(\chi) \\
         \cos^{N-1}(\chi)  & \hspace{-3mm} 1 
     \end{pmatrix}, 
     \label{trace_squeezed}
\end{equation}
as proved in Appendix~\ref{sec:appendix_trace_squeezed}. The local coherence and entanglement of the state are both governed by the squeezing parameter $\chi$: varying $\chi$ continuously transforms the state from a product state with maximal coherence ($\chi=0$) to a maximally entangled state with no coherence ($\chi=\pi/2$). Indeed, at $\chi=\pi/2$, one obtains the maximally entangled state $|\psi_{\rm sq}(\pi/2)\rangle=e^{-i\pi(N^2/8-1/4)}(|a\rangle^{\otimes N}-i|b\rangle^{\otimes N})/\sqrt{2}$, where $|a\rangle=|0\rangle -e^{iN\pi/2}|1\rangle$ and $|b\rangle=|0\rangle +e^{iN\pi/2}|1\rangle$ are orthogonal states.
Conversely, the local temperature is fixed, regardless of the value of $\chi$. Notably, for $\chi \neq n \pi$ the reduced density operator in Eq.~\eqref{trace_squeezed} approaches the maximally mixed state in the large-$N$ limit.

\section{Results}\label{sec:results}

We now present the results of our analysis, focusing on how specific properties of the bath and of the thermometer initial state influence the QFI~\cite{note_wolfram}. The properties under investigation are:
(i) the inverse temperature of the bath $\beta$, (ii) the local inverse temperature $\mu$ of the thermometers as defined in Eq.~\eqref{eq:local_inverse_temperature_mu}, (iii) the local coherence of the thermometers and (iv) the quantum correlations among the $N$ thermometers. The analysis on quantum correlations and the interplay between them and the other properties represent the main novel contribution of this work.

Each physical property $\mathcal{P}$ listed above is controlled in the input states $\hat{\rho}_{\rm in}$ by some set of parameters (see Sec.~\ref{sec:input_states}) that we collectively denote as $\varphi$. We denote by $\varphi_{\rm max}$ ($\varphi_{\rm min}$) the set of parameters that maximizes (minimizes) the property $\mathcal{P}$ in $\hat{\rho}_{\rm in}$. Furthermore, we introduce the shorthand notation  $\mathcal{F}_Q^*\big(\hat{\rho}_{\rm in}(\varphi) \big):=\max_t\mathcal{F}_Q\big((\Phi_\beta^{[t]})^{\otimes N}[\hat{\rho}_{\rm in}(\varphi)] \big)$. We then define
\begin{equation}
    G_\mathcal{P} (\hat{\rho}_{\text{in}}) = \frac{ \mathcal{F}_Q^*\big(\hat{\rho}_{\text{in}}(\varphi_{\textrm{max}})\big)}{\mathcal{F}_Q^*\big(\hat{\rho}_{\text{in}}(\varphi_{\textrm{min}})\big)}.
    \label{eq:gain}
    \end{equation}
This quantity determines a potential sensitivity \textit{gain} offered by $\mathcal{P}$ by comparing the maximum achievable QFI when the property $\mathcal{P}$ in the input state is maximized to the case where it is minimized or absent. $G_\mathcal{P} (\hat{\rho}_{\text{in}})>1$ indicates a positive contribution of the property $\mathcal{P}$ to the QFI, signifying an enhancement in sensitivity. If $G_\mathcal{P} (\hat{\rho}_{\text{in}})=1$, then $\mathcal{P}$ provides no advantage. This situation includes, for instance, cases where the maximal QFI is achieved asymptotically: any difference determined by the properties of the input state is lost, in such case, due to thermalization. Conversely, a value of $G_\mathcal{P} (\hat{\rho}_{\text{in}})$ in the range $0\leq G_\mathcal{P} (\hat{\rho}_{\text{in}})<1$ implies that the presence of $\mathcal{P}$ is actually detrimental to the QFI and to the sensitivity.

In order to be able to manipulate single properties of various states, we define a new $N$-qubit state where all correlations contained in $\hat{\rho}$ have been removed:  
\begin{equation}
    \hat{\rho}^\nc=\left(\tr_{N-1}[\hat{\rho}]\right)^{\otimes N}.
    \label{eq:rho_no_correlations}
\end{equation}

In the final part of the \textit{Results} section, we examine how the QFI of some correlated input states scales with the number $N$ of thermometers. We also compare the scaling behaviors obtained from these states with the SQL and the upper bound discussed in Sec.~\ref{sec:qfi_bound}. 

\subsection{The role of bath and thermometers' temperatures}
\label{sec:temperature_role}

\begin{figure}[t]
    \centering
        \includegraphics[width=0.95\linewidth]{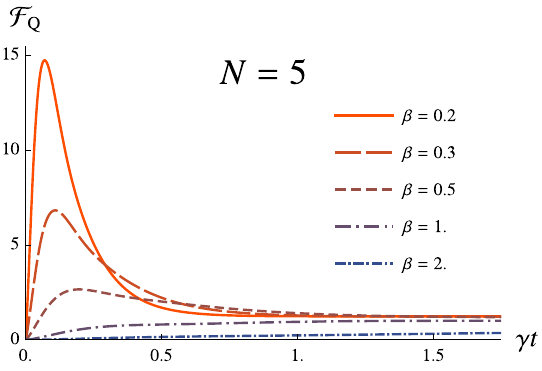}
        \vspace{2mm}
        $\thmix(\eta=0.75,\mu=1.5)$
        \label{fig:differentbeta_tm}
    \caption{
    \justifying QFI of the evolved state of an ensemble of 5 qubits initialized in the $\thmix$ state. Different colors correspond to different bath temperatures. Similar results can be obtained for all the analyzed input states, both correlated and uncorrelated. A more consistent improvement in the precision of the estimate of $\beta$ can be obtained for hotter baths.}
    \label{fig:effect_of_beta}
\end{figure}

\begin{figure}[t]
    \centering
    \includegraphics[width=
    0.95\linewidth]{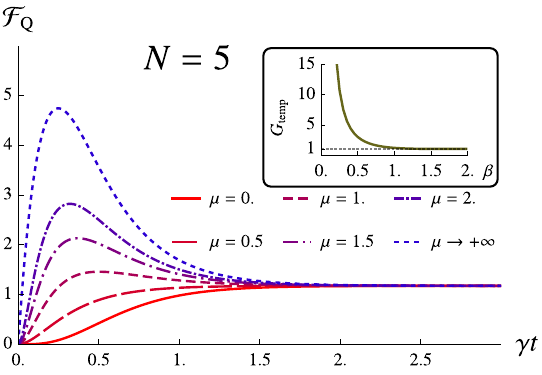}
    \vspace{2mm}
    $\thmix^\nc(\mu), \: \beta = 0.5$.
    \caption{\justifying QFI of the evolved thermometers' ensemble state, initialized in the thermal state. Different colors correspond to different values of the local thermometers' inverse temperature $\mu$. Inset: gain $G_\text{temp}(\thmix^\nc)$ due to a low local temperature of the input state, defined in Eq.~\eqref{eq:Gtemp_thermal}, as a function of $\beta$. A lower local temperature of the qubits proves to be beneficial for the QFI, especially when estimating the inverse temperature of hot thermal baths. The SQL is recovered in the limit $\mu \to \infty$ as $\thmix^\nc$ approaches the ground state.}
    \label{fig:variations_with_mu}
\end{figure}

In Sec.~\ref{sec:qfi_bound}, we have shown how the sensitivity limits provided by the SQL and the upper bound on the channel QFI depends on $\beta$. Both indicate an enhanced sensitivity when probing a hot bath (small $\beta$) as compared to a cold one (large $\beta$). This behavior, which has been observed in a variety of thermometric systems~\cite{shu_2020,mukherjee2019enhanced,correa_2017,zhang2024enhancinglowtemperaturequantumthermometry,Potts2019fundamentallimits,ullah_2023}---including the single-qubit thermometer~\cite{Fraz_o_2024}---is recovered in the analysis of the multiqubit correlated states introduced in Sec. \ref{sec:input_states}. As a representative example, Fig.~\ref{fig:effect_of_beta} illustrates the behavior of the QFI associated with a five-qubit thermometer prepared in a thermal-mixtured GHZ state. The QFI exhibits an earlier and more marked peak during the transient for small values of $\beta$, similarly to what is seen in Fig.~\ref{fig:bound_figure}. Moreover, the time required for the QFI to converge to its asymptotic value decreases as the bath temperature increases. This result is consistent with the fact that, according to Eq.~\eqref{eq:lambda}, the thermalization rate $\lambda$ increases for smaller values of $\beta$: an higher bath temperature entails a faster relaxation of each thermometer to thermal equilibrium.

We next compare the influence of the bath temperature with the effect of a variation in the thermometers' local temperature. Since $\mu$ controls more than one input resource, we remove correlations from $\thmix(\eta,\mu)$ to isolate the effect of local temperature, postponing an analysis of the interplay between different resources to a later section. Following Eq.~\eqref{eq:rho_no_correlations}, we get the input state
\begin{equation}
    \thmix^\nc (\mu) = \thermal^{\otimes N}(\mu).
\end{equation}

In Fig.~\ref{fig:variations_with_mu}, we plot the QFI obtained by preparing 5 thermometers in the $\thmix^\nc$ state, fixing the value of the inverse bath temperature at $\beta=0.5$ and allowing $\mu$ to vary. Our results show that reducing the local temperature of the thermometers' input state (increasing $\mu$) has a positive effect on the QFI, similar to that induced by increasing the bath temperature. 
Consequently, for a fixed value of $\beta$, the optimal performance is attained in the limit $\mu \to +\infty$, which corresponds to the SQL associated with the given temperature of the bath. In this regime, indeed, the state $\thmix^\nc$ approaches the ground state $\ket{0}^{\otimes N}$.

The improvement in sensitivity due to the local temperature of the input state can be quantified by the gain
\begin{equation}
    G_\text{temp}(\thmix^\nc) = \frac{\mathcal{F}_Q^*\big(\ket{0}^{\otimes N}\big)}{\mathcal{F}_Q^*(\id_{2^N}/{2^N})},
    \label{eq:Gtemp_thermal}
\end{equation}
defined according to Eq.~\eqref{eq:gain}. This quantity compares the performance achieved in the limit $\mu\to+\infty$ with that at $\mu=0$. As shown in the inset of Fig.~\ref{fig:variations_with_mu}, cold input states are always the most favorable, since $G_\text{temp} \geq 1$  for all values of $\beta$. However, a substantial gain is only achieved at low values of $\beta$---where a transient peak exists---while $G_{\rm temp}$ decrease to 1 for high values of $\beta$---where the maximum QFI is achieved asymptotically regardless of the input state.

\subsection{The role of quantum coherence}
\label{sec:coherence_role}

Some of our input states are characterized by a fixed local temperature but exhibit varying degrees of local coherence. To distinguish the effects of this resource from those of quantum correlations, it is therefore necessary to investigate its influence on the QFI. The amount of coherence can be adjusted in a squeezed-spin input state $|\psi_{\rm sq}(\chi)\rangle$ by tuning the squeezing parameter; however, $\chi$ also controls entanglement. For this reason, following Eq.~\eqref{eq:rho_no_correlations}, we remove initial correlations, getting
\begin{equation}
    \squeez^\nc(\chi)=\frac{1}{2^N}
     \begin{pmatrix}
         1 & \hspace{-3mm} \cos^{N-1}(\chi) \\
         \cos^{N-1}(\chi)  & \hspace{-3mm} 1 
     \end{pmatrix}^{\otimes N}\label{eq:squeezed_no_correlations}
\end{equation}
as the thermometer input state. 
This state attains maximal local coherence for $\chi=0$, when off-diagonal elements are maximized, whereas it contains no coherence for $\chi = \pi/2$, when these vanish. 

\begin{figure}[t]
    \centering
\includegraphics[width=0.95\linewidth]{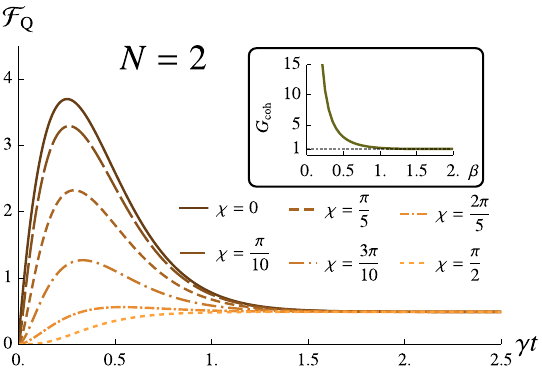}
\vspace{2mm}
{$\squeez^\nc(\chi)$, $\beta=0.3$. }
    \caption{QFI of the evolved thermometers' ensemble state, initialized in the tensor product of the reduced state of the squeezed spin state for two qubits. Different colors correspond to different values of the local coherence, regulated by $\chi$. The inverse bath temperature is set at $\beta=0.3$. Inset: gain $G_\text{coh}(\squeez^\nc)$ due to the presence of local coherence in the input state, defined in Eq.~\eqref{eq:Gcoh_squeezed_nc}, as a function of $\beta$. Local coherence improves the precision on the estimate of $\beta$, especially at high values of the bath temperature.}
    \label{fig:sqp_coherence}
\end{figure}

Results obtained from numerical simulations are shown in Fig.~\ref{fig:sqp_coherence}, where the QFI is depicted as a function of time for different values of $\chi\in [0, \pi/2]$, keeping the inverse bath temperature fixed at $\beta = 0.3$. By increasing $\chi$, or, equivalently, by decreasing the local coherence, we observe a consequent reduction of the QFI during the transient. Consistently with existing literature~\cite{Fraz_o_2024}, this demonstrates the beneficial role played by quantum coherence in temperature estimation. 

The same conclusion can be drawn by examining the inset of Fig.~\ref{fig:sqp_coherence}. Here, we plot the gain 
\begin{equation}
    G_\text{coh}(\squeez^\nc) = \frac{\mathcal{F}_Q^*\big(\ket{+}^{\otimes N}\big)}{\mathcal{F}_Q^*(\id_{2^N}/2^N)}\new{,}
    \label{eq:Gcoh_squeezed_nc}
\end{equation}
which compares the performance achieved with $\chi=0$ to $\chi=\pi/2$. The gain curve lies above 1 for every value of the bath temperature, but substantial gains are only achieved fo low values of $\beta$. Once again, this behavior highlights that thermometry of hot baths is comparatively easier than operating at low temperatures.

\subsection{The role of quantum correlations}
\label{sec:correlations_role}

\begin{figure}[t]
    \centering
        \includegraphics[width=0.45
        \textwidth]{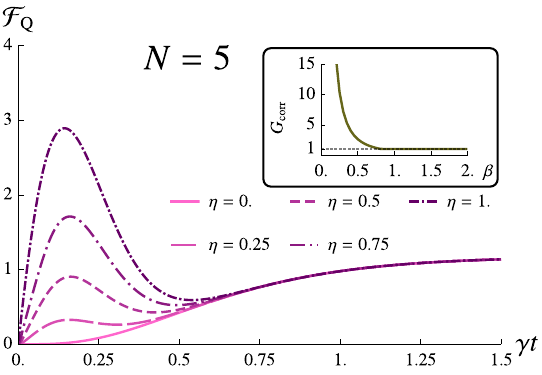}
        \label{fig:subfig1}
        \vspace{2mm}
        {$\idmix (\eta)$, $\beta = 0.5$}
    \caption{\justifying QFI of the evolved state of a five-qubit thermometer initialized in the identity-mixtured GHZ state $\idmix$ as a function of time. Different colors correspond to  different values of the correlation parameter $\eta \in [0,1]$. The inverse bath temperature is set at $\beta=0.5$. Inset: gain $G_\text{corr}(\idmix)$ due to the presence of quantum correlation in the input state, defined in Eq.~\eqref{eq:gain_cor_idmix}, as a function of $\beta$. When the reduced states of each qubit are held fixed, the presence of initial quantum correlations enhances the QFI during the transient, especially when probing hot baths.  
    Accordingly, quantum correlations can serve as an alternative to the local inverse temperature in enhancing the precision of the temperature estimation, as illustrated in Fig.~\ref{fig:variations_with_mu}.}
    \label{fig:1mix_eta}  
\end{figure}

We now come to analyze the effect of quantum correlations in the input state of the thermometers. We start by considering the identity-mixtured GHZ input state of Eq.~\eqref{eq:ghz_mix} , which can be obtained from the thermal-mixtured GHZ state of Sec.~\ref{sec:temperature_role} by setting $\mu=0$. Interestingly, this state has vanishing inverse local temperature and coherence, and thus represents the most inefficient case from the perspective of the other input resources. Therefore, any aspect of its thermometric performance can be attributed entirely to the presence of quantum correlations, whose amount can be adjusted by tuning the parameter $\eta$.

Fig.~\ref{fig:1mix_eta} shows the QFI of the evolved state for different values of $\eta$. The inverse bath temperature is set at $\beta=0.5$. As $\eta$ increases, a clear progression of dynamical features is observed: $(i)$ the emergence of a peak during the transient regime at $\eta=0.25$; $(ii)$ the peak exceeding the QFI asymptotic value at $\eta=0.75$, and $(iii)$ attaining its maximum at $\eta=1$, corresponding to the case in which the initial state becomes a GHZ state. Notably, when other input resources are minimized, the use of correlated probes can still secure an advantage in sensitivity compared to thermal equilibrium. Moreover, a similar behavior is obtained for different values of $\beta$ and for input states with a non-vanishing inverse local temperature (i.e., for thermally-mixtured states with $\mu \neq 0$). These results unambiguously establish quantum correlations as a valuable resource for nonequilibrium quantum thermometry, alongside the local resources well characterized in previous works. We also notice that the amount of correlations required to generate a transient peak that exceeds the QFI asymptotic value increases with $\beta$, in agreement with the results discussed in Sec.~\ref{sec:temperature_role}.  

In the inset of the same figure, we plot the gain
\begin{equation}
    G_\text{corr} (\idmix) = \frac{\mathcal{F}_Q^*\big(\ket{\text{GHZ}}\big)}{\mathcal{F}_Q^*\big( \id_{2^N}/2^N\big)},
    \label{eq:gain_cor_idmix}
\end{equation}
which compares the performances achieved with $\eta=1$ and $\eta=0$, as a function of $\beta$. This quantity shows a behavior analogous to that of $G_{\rm temp}$ and $G_{\rm coh}$ (see Secs.~\ref{sec:temperature_role} and \ref{sec:coherence_role}, respectively). We find $G_\text{corr}\geq 1$, with larger values for smaller $\beta$, indicating that quantum correlations are particularly useful when probing hot baths. More complete details regarding this type of gain will be provided in the following sections.

We provide further evidence of a correlation-induced sensitivity gain in Fig.~\ref{fig:sq_minus_kronprod}. To quantify the contribution of quantum correlations, we compare the QFI obtained from the squeezed input state, Eq.~\eqref{eq:squeezed_state}, with that achieved after removing all its initial correlations, as in Eq.~\eqref{eq:squeezed_no_correlations}. The difference between the two QFIs is plotted in Fig.~\ref{fig:sq_minus_kronprod} as a function of time $\tau = \gamma t$, for different values of the squeezing parameter $\chi$. This quantity increases as $\chi$ approaches $\pi/2$, when the squeezed spin state possesses a larger amount of quantum correlations, and it is maximized when the input state becomes maximally entangled, for $\chi=\pi/2$, as one would expect for an enhancement arising from quantum correlations. 

If a general conclusion is to be drawn from our results up to this point, it would be that, among states sharing the same single-qubit partial trace, those endowed with the largest amount of quantum correlations consistently yield the best precisions. In the next two subsections, we further examine quantum correlations and their interplay with local temperature and local coherence in specific classes of states.

\subsection{Interplay between quantum correlations and local temperature}
\label{sec:intermplay_correlations_localtemperature}

\begin{figure}[t]
    \centering
    \includegraphics[width=0.95\linewidth]{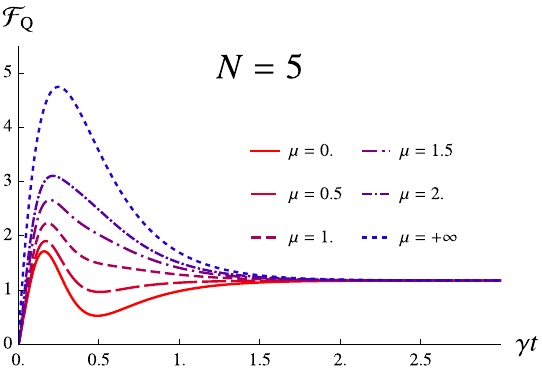}
    \label{subfig:thmix_loctemp}
    \vspace{2mm}
    $\thmix(\eta=0.75,\mu), \: \beta = 0.5$.
    \caption{QFI of the evolved thermometers' ensemble state, initialized in the $\thmix$ state of 5 qubits. Different colors correspond to different values of the local thermometers' inverse temperature $\mu$. A lower local temperature of the qubits proves to be beneficial for the QFI. By comparing the figure with Fig.~\ref{fig:variations_with_mu}, we observe that the presence of quantum correlations improves the thermometers' sensitivity, especially when their local temperatures is high.}
\label{fig:thmix_different_local_temperatures}
\end{figure}

\begin{figure}[t]
    \centering
    \includegraphics[width=0.975\columnwidth]{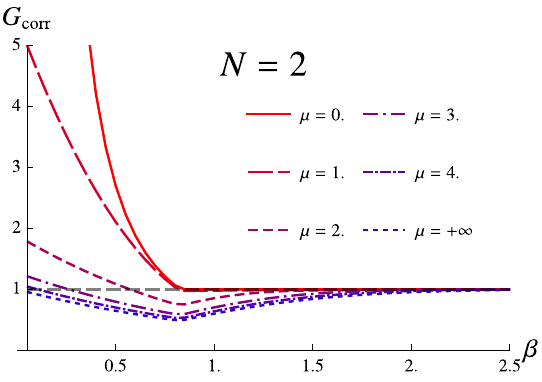}
    \vspace{2mm}
    $\thmix (\mu)$
    \caption{\justifying Gain $G_{\rm corr}$ of the five-qubit $\thmix$ state, defined in Eq.~\eqref{eq:gain_cor_thmix}, as a function of $\beta$. Different colors correspond to different values of $\mu$ characterizing the initial state. As $\mu$ increases, both the local temperature of the thermometers and the correlation between them decrease. Since the curve comparing the GHZ with the ground initial states is always non-positive, we can conclude that the maximally entangled state does not exceed the SQL. Remarkably, however, for extremely hot baths ($\beta \to 0$), the quantity $G_\text{corr}$ goes to 1: in this regime, quantum correlations can compensate for an otherwise disadvantageous high local temperature, allowing for performances close to the SQL.}
    \label{fig:Gcorr_thmix_vs_beta}
\end{figure}

For a multiqubit input state, a full characterization of the thermometric performance is possible only when the interplay among local and nonlocal input resources is properly taken into account. In this section, we analyze the interplay between local temperature and quantum correlations.

As a first step in this analysis, we revisit the case study presented in Fig.~\ref{fig:variations_with_mu}, now restoring quantum correlations in the input state. In Fig.~\ref{fig:thmix_different_local_temperatures} we consider a thermally-mixtured GHZ state whose mixing parameter is fixed to $\eta=0.75$. In this case too, decreasing the local temperature of the initial state of the thermometers increases their sensitivity during the transient. Moreover, comparing with Fig.~\ref{fig:variations_with_mu}, we find evidence of a sensitivity enhancement induced by quantum correlations. This is more pronounced at higher local temperatures: in particular, small transient maxima now appear even at very low $\mu$ values. However, none of the entangled input states surpass the SQL, which corresponds to the $\mu\to+\infty$ curve in the figure. That is, thermometer preparation in the $\ket{0}^{\otimes N}$ state entails the highest QFI peak. While highlighting undeniable advantages for quantum thermometry, our analysis thus also reveals some of the limitations associated with the use of entangled states, particularly when compared to certain optimal product states.

This coexistence of advantages and limitations is further addressed and clarified by the plot in Fig.~\ref{fig:Gcorr_thmix_vs_beta}, where we evaluate the gain
\begin{equation}
    G_\text{corr} (\thmix) = \frac{\mathcal{F}_Q^*\big(\ket{\text{GHZ}}\big)}{\mathcal{F}_Q^*\big( \thermal^{\otimes N}(\mu)\big)}.
    \label{eq:gain_cor_thmix}
\end{equation}
This extends the analysis of Sec.~\ref{sec:correlations_role} (cfr. Eq.~\eqref{eq:gain_cor_idmix}) by comparing the performance of the GHZ state---characterized by a vanishing inverse local temperature and local coherence---with that of a thermal product state with a variable local temperature. Within the family of thermally-mixtured states defined by Eq.~\eqref{eq:thermal_ghz_mix}, these two states respectively exhibit the maximum and minimum amount of quantum correlations. In Fig.~\ref{fig:Gcorr_thmix_vs_beta}, $G_\text{corr}$ is plotted as a function of $\beta$ for different values of $\mu$. The common behavior exhibited by the different gain curves allows us to identify two distinct $\beta$ regimes. In the high-$\beta$ regime, where the gain increases with $\beta$, the GHZ state does not generate a transient peak; consequently, $G_\text{corr}$ effectively compares the peak of the product state with the QFI asymptotic value. By contrast, in the more relevant low-$\beta$ regime---where the gain decreases with $\beta$---distinct transient peaks are generated by both states considered in Eq.~\eqref{eq:gain_cor_thmix}. Here, we can further distinguish three distinct regimes depending on the value of $\mu$. For small $\mu$ ($\mu\leq 1$), the gain never falls below unity. This behavior, analogous to that observed in Fig.~\ref{fig:1mix_eta}, shows that hot thermometers which lack quantum correlations can never outperform the GHZ state. However, as $\mu$ is increased to moderately colder temperatures ($2\leq\mu\leq 3$), an expanding region of $\beta$ values emerges in which the gain falls below unity. Only extremely cold thermometers ($\mu\geq4$) are found to outperform the GHZ state unconditionally, as the gain curves lie entirely below unity. It should be noted, however, that although these states do not require quantum correlations to achieve optimal performance, their practical realization is hindered by the significant thermodynamical effort required to cool the uncorrelated qubit probes. We also see that the sensitivity advantage of these states over the GHZ state tends to vanish as $\beta \to 0$, since $G_\text{corr}\to 1$. This means that cold uncorrelated and maximally entangled states perform equally well when probing extremely hot baths, both achieving the optimal thermometric sensitivity in this temperature regime.

\subsection{Interplay between quantum correlations and local coherence}
\label{sec:interplay_correlations_coherence}

\begin{figure*}[t]
    \centering
    
    \begin{minipage}[t]{0.5\textwidth}
    \vspace{7mm}
        \centering
        \subfloat[$\squeezvec{\chi}$, $\beta=0.5$]{%
            \includegraphics[width=\linewidth]{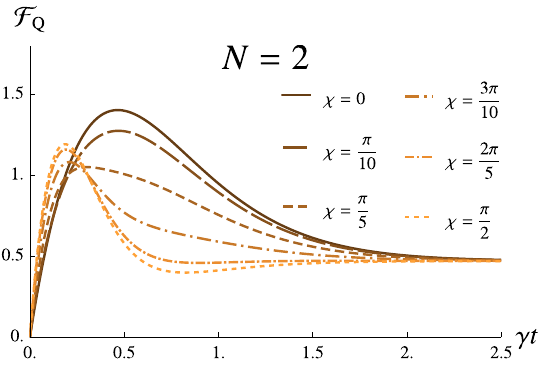} 
            \label{subfig:qfi_sq_2d}
        }
    \end{minipage}
    \hfill
    \begin{minipage}[t]{0.48\textwidth}
        \centering
        \subfloat[QFI 3D]{%
    \includegraphics[width=0.48\linewidth]{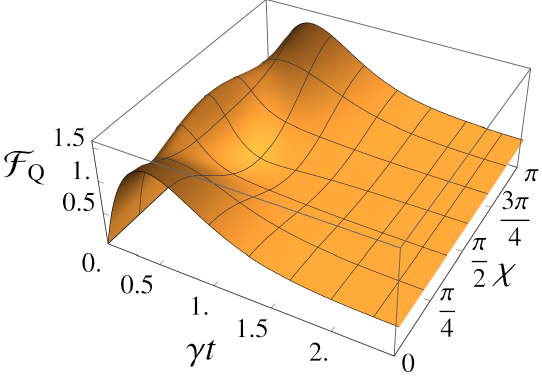} 
            \label{subfig:qfi_sq_3d}
        }
        \hfill
        \subfloat[Gain $G_\text{corr}$]{%
            \includegraphics[width=0.48\linewidth]{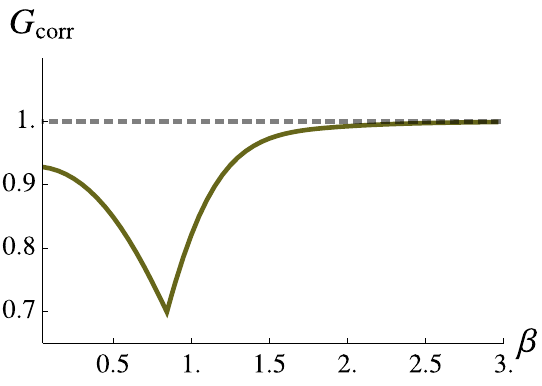} 
            \label{subfig:gain_sq}
        }
        \\
        \subfloat[Negativity of entanglement]{%
            \includegraphics[width=0.48\linewidth]{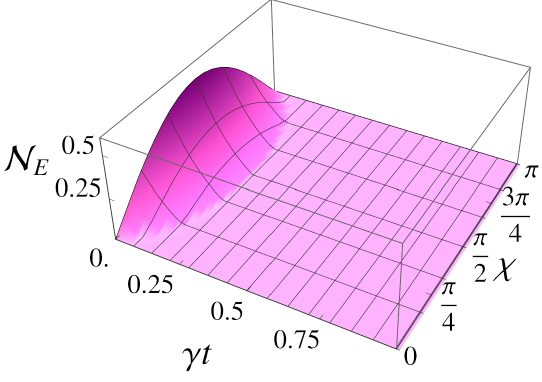}
            \label{subfig:negativity_sq_3d}
        }
        \hfill
        \subfloat[Local coherence]{%
            \includegraphics[width=0.48\linewidth]{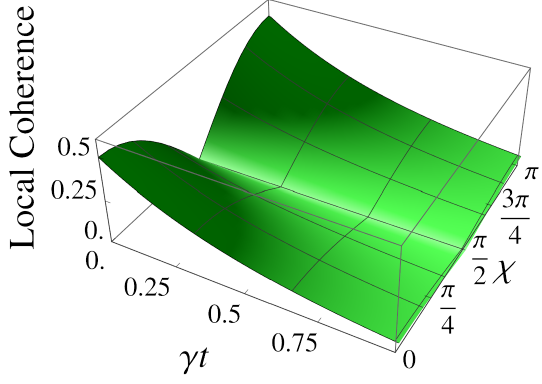}
            \label{subfig:coherence_sq_3d}
        }
 \end{minipage}

    \caption{
        \justifying Panels (a)-(b): QFI of a two-qubit state initialized in the squeezed spin state of Eq.~\eqref{eq:squeezed_state}. In panel (a), the QFI is plotted as a function of the normalized time $\gamma t$, and different curves correspond to different values of the squeezing parameter $\chi$. Panel (b) shows the QFI as a two-variable function of $\gamma t$ and $\chi$. The QFI, being periodic in $\chi$ with period $\pi$, is plotted on the interval $\chi\in[0,\pi]$. The inverse temperature of the bath is fixed at $\beta = 0.5$. 
        Panel (c): Gain $G_\text{corr}(\ket{\psi_{\text{sq}}})$ due to correlations, defined in Eq.~\eqref{eq:Gcorr_squeezed}, as a function of $\beta$. For small values of $\beta$. i.e. hot baths, the lack of local coherence in the initial thermometers state can be compensated by the presence of quantum correlations among the qubits.
        Panels (d)-(e): Negativity of entanglement (cfr. Eq.~\eqref{eq:negativity}) and local coherence, respectively, of the two-qubit state as functions of time and of the parameter $\chi$. As for the QFI, these two quantities are periodic in $\chi$ with period $\pi$ and here plotted on the interval $\chi\in[0,\pi]$. The inverse temperature of the bath is fixed at $\beta = 0.5$. For this input state, enhancements of the transient QFI can be ascribed to the presence of local coherence and/or quantum correlations. 
    }
    \label{fig:squeezed_beautiful_plots}
\end{figure*}

Our aim in this section is to clarify how the QFI is affected by the interplay between quantum correlations and local coherence. For this analysis, we select those correlated input states whose control parameters do not affect the state's local temperature: the squeezed spin state, Eq.~\eqref{eq:squeezed_state}, and the $k-$GHZ superposition state, Eq.~\eqref{eq:k_GHZ_superposition}, with $|k\rangle=|+\rangle$. Working with two-qubit states, we can precisely quantify the amount of entanglement in the initial state and during evolution. To this purpose, we use the \textit{negativity of entanglement} $\mathcal{N}_E$~\cite{Horodecki_2009}: 
\begin{equation}
    \mathcal{N}_E \equiv \abs{\sum_{\lambda_j < 0} \lambda_j},
    \label{eq:negativity}
\end{equation}
where $\lambda_j$ are obtained by applying partial transposition on the thermometer state and then extracting its negative eigenvalues. $\mathcal{N}_E$ vanishes on separable states; for two-qubit states, one has $0\leq\mathcal{N}_E\leq 1/2$ and $\mathcal{N}_E=1/2$ if and only if the state is maximally entangled.

First, we consider the  squeezed spin input state. The QFI of the evolved state is shown in Fig.~\ref{subfig:qfi_sq_2d} as a function of time and for different values of $\chi$, the squeezing parameter of the input state. As a side note, we observe that the QFI displays a symmetry in the squeezing parameter such that the input states $\squeezvec{\chi}$ and $\squeezvec{\pi-\chi}$ lead to the same values of the QFI over time. Transient peaks of variable magnitude are observed across all curves. A comparison with Fig.~\ref{fig:sqp_coherence} reveals that, while peaks around $\chi = 0$ and $\chi = \pi$ persist even when correlations are removed from the initial state---owing to a high degree of coherence---peaks around $\chi = \pi/2$ emerge only in the presence of quantum correlations. Notably, while the largest peak occurs at $\chi=0, \pi$, corresponding to maximal coherence and the absence of entanglement, a secondary local maximum in the peak height is obtained for $\chi=\pi/2$, i.e., when the thermometer is prepared in a maximally entangled state with no local coherence. This behavior is confirmed by a three-dimensional plot of the QFI as a function of both $t$ and $\chi$ (Fig.~\ref{subfig:qfi_sq_3d}). In Fig.~\ref{subfig:gain_sq}, we plot the gain
\begin{equation}
    G_\text{corr}(\ket{\psi_\text{sq}}) = \frac{\mathcal{F}_Q^*\big(\squeezvec{\pi/2}\big)}{\mathcal{F}_Q^*\big(\ket{+}^{\otimes N} \big)}
    \label{eq:Gcorr_squeezed}
\end{equation}
as a function of $\beta$, for a quantitative comparison of the optimal sensitivities achieved by $\chi=\pi/2$ and $\chi=0$. As the curve stays consistently below unity, the plot confirms that thermometers with high local coherence do outperform highly entangled input states despite lacking quantum correlations, similarly to those characterized by low local temperature. In the high-$\beta$ regime, where the gain increases with $\beta$, only the $\ket{+}^{\otimes N}$ product state exhibits a transient peak, and $G_\text{corr}\to 1$ as the height of this peak approximates the QFI asymptotic value. Instead, in the more relevant low-$\beta$ regime, both input states give rise to a transient peak. In this case, the gain decreases with $\beta$ and the sensitivity gap also progressively narrows as $\beta \to 0$. Therefore, highly coherent and highly entangled states deliver almost identical performances when probing extremely hot baths.

As anticipated above, for a two-qubit thermometer we can track the temporal evolution of both local coherence and entanglement. We provide three-dimensional plots of these two quantities as functions of both time and $\chi$ in the last two panels of Fig.~\ref{fig:squeezed_beautiful_plots}. Fig.~\ref{subfig:negativity_sq_3d} shows the negativity of entanglement, $\mathcal{N}_E$, defined in Eq.~\eqref{eq:negativity}. Fig.~\ref{subfig:coherence_sq_3d} displays the state local coherence, derived from the module of the off-diagonal elements of the evolved density matrix. A comparison with Figs.~\ref{subfig:qfi_sq_2d} and~\ref{subfig:qfi_sq_3d} reveals several interesting features. In particular, the early and narrow peaks observed in the QFI curves at values of $\chi$ close to $\pi/2$ appear to correlate with the rapid temporal decay of the negativity of entanglement during the early stages of the thermalization process. By contrast, the curves associated with $\chi$ values around $0$ and $\pi$ exhibit later and larger transient peaks, which are consistent with the slower decay of local coherence. The temporal behavior of these quantities therefore supports our intuition that distinct quantum resources are responsible for different sensitivity peaks in the QFI curves.

\begin{figure*}[t!]
    \centering

    \begin{minipage}[t]{0.48\textwidth}
    \vspace{7mm}
        \centering
        \subfloat[$\plusghzvec{\alpha}$, $\beta=0.5$]{%
            \includegraphics[width=\linewidth]{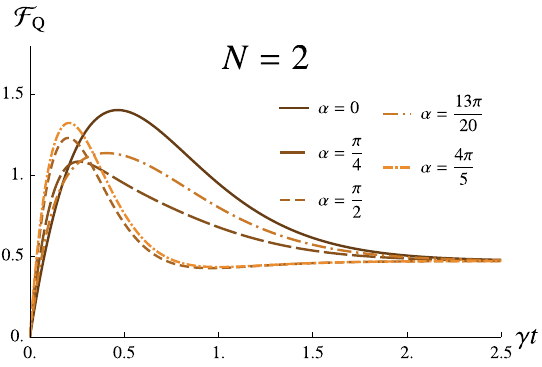} 
            \label{subfig:qfi_plusghz_2d}
        }
    \end{minipage}
    \hfill
    \begin{minipage}[t]{0.48\textwidth}
        \centering

        \subfloat[QFI 3D]{%
            \includegraphics[width=0.48\linewidth]{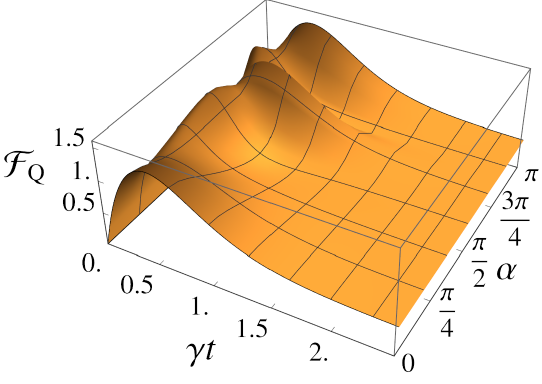} 
            \label{subfig:qfi_plusghz_3d}
        }
        \hfill
        \subfloat[Gain $G_\text{corr}$]{%
            \includegraphics[width=0.48\linewidth]{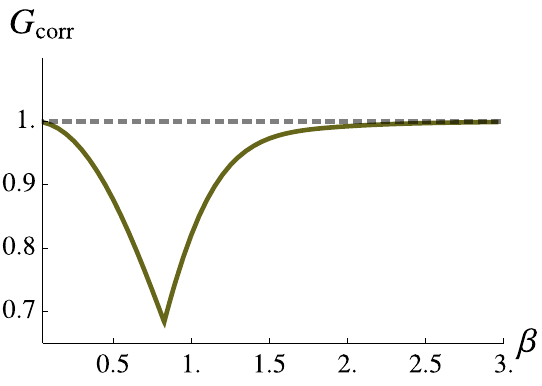} 
            \label{subfig:qfi_plusghz_gain}
        }
        \\   
        \subfloat[Negativity of entanglement]{%
            \includegraphics[width=0.48\linewidth]{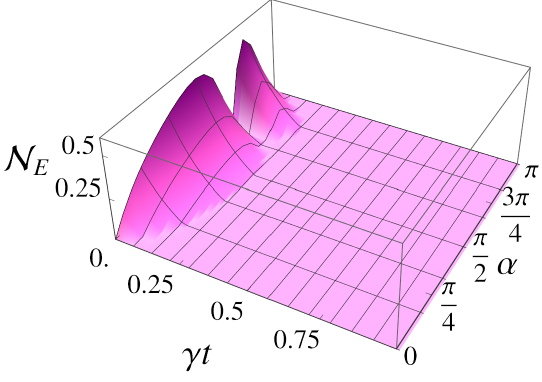}
            \label{subfig:qfi_plusghz_neg}
        }
        \hfill
        \subfloat[Local coherence]{%
            \includegraphics[width=0.48\linewidth]{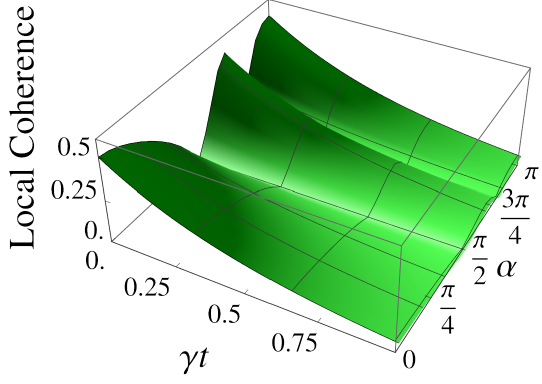}
            \label{subfig:qfi_plusghz_coherence}
        }

    \end{minipage}

    \caption{
        \justifying Panels (a)-(b): QFI of a two-qubit state initialized in the $k$-GHZ superposition state of Eq.~\eqref{eq:k_GHZ_superposition} with $\ket{k} = \ket{+}$. In panel (a), the QFI is plotted as a function of the normalized time $\gamma t$, and different curves correspond to different values of the parameter $\alpha$. Panel (b) shows the QFI as a two-variable function of $\gamma t$ and $\alpha$ and the parameter $\alpha$. The QFI, being periodic in $\alpha$ with period $\pi$, is plotted on the interval $\alpha \in[0,\pi]$. The inverse temperature of the bath is fixed at $\beta = 0.5$.  
        Panel (c): Gain $G_\text{corr} (\ket{\psi_{\text{+-s}}})$, defined in Eq.~\eqref{eq:Gcorr_plusghz}, as a function of $\beta$. For small values of $\beta$. i.e. hot baths, the lack of local coherence in the initial thermometers state can be compensated by the presence of quantum correlations among the qubits. In particular, for $\beta \to 0$ the performance of the GHZ state appears to match the SQL.
        Panels (d)-(e): Negativity of entanglement (cfr. Eq.~\eqref{eq:negativity}) and local coherence, respectively, of the two-qubit state as functions of time and of the parameter $\alpha$. As for the QFI, these two quantities are periodic in $\alpha$ with period $\pi$ and here plotted on the interval $\alpha \in[0,\pi]$. The inverse temperature of the bath is fixed at $\beta = 0.5$. For this input state, enhancements of the transient QFI can be ascribed to the presence of local coherence and/or quantum correlations.
    }
    \label{fig:plusghz_beautiful_plots}
\end{figure*} 

Analogous results are obtained when employing the $k$-GHZ superposition state, suggesting that our conclusions are general and not tied to a particular choice of input state. Fixing $|k\rangle=|+\rangle$, the state parameter $\alpha$ governs the interplay between initial local coherence and entanglement (as $\chi$ for the $\squeezvec{\chi}$ state). Fig.~\ref{fig:plusghz_beautiful_plots} presents the results obtained with this input state; the panel organization is the same as in Fig.~\ref{fig:squeezed_beautiful_plots}. The QFI landscape appears even richer and more complex than in the previous case: Fig.~\ref{subfig:qfi_plusghz_3d} shows five local maxima in the height of the transient peaks. With the aids of Figs.~\ref{subfig:qfi_plusghz_neg} and \ref{subfig:qfi_plusghz_coherence}, these maxima can be attributed to one of two quantum resources, namely quantum correlations or local coherence. Two of the five local maxima, which occur at early times, are generated by maximally entangled states, specifically the GHZ state ($\alpha=\pi/2$) and the Bell state $[|01\rangle+|10\rangle]/\sqrt{2}$ ($\alpha=3\pi/4$). The remaining three maxima occur at later times and are produced by states exhibiting maximal coherence, corresponding to $\alpha=0$, $\alpha=\arcsin{[1/\sqrt{3}]}+\pi/2$ and $\alpha=\pi$. In Fig.~\ref{subfig:qfi_plusghz_gain}, we plot the gain
\begin{equation}
    G_\text{corr}(\ket{\psi_\text{+-s}}) = \frac{\mathcal{F}_Q^*\big(\ket{\text{GHZ}}\big)}{\mathcal{F}_Q^*\big(\ket{+}^{\otimes N} \big)}
    \label{eq:Gcorr_plusghz}
\end{equation} 
as a function of $\beta$, in order to compare the optimal sensitivities achieved by $\alpha=\pi/2$ and $\alpha=0$. The observed behavior closely resembles that shown in Fig.~\ref{subfig:gain_sq}, and therefore support analogous conclusions.

Finally, we observe that by considering the tensor products of the single-qubit reduced states of the two-qubit states $\squeezvec{\chi}$ and $\plusghzvec{\alpha}$---namely, $\plusghz^\nc (\alpha)$ and $\squeez^\nc(\chi)$---we observe that, by choosing $\chi$ and $\alpha$ such that all the local coherences vanish, both constructions yield the maximally mixed input state $\id_4/4$. In agreement with the findings of Secs.~\ref{sec:temperature_role}, \ref{sec:coherence_role} and \ref{sec:correlations_role}, the maximally mixed state exhibits the worst metrological performance, as it lacks quantum correlations, possesses no local coherence, and has infinite local temperature.

\vspace{2mm}
\subsection{Scaling of the maximal QFI with $N$}
\vspace{-2mm}

\begin{figure}[t]
    \centering
    \subfloat[$\beta=0.1$]{\includegraphics[width=0.87\linewidth]{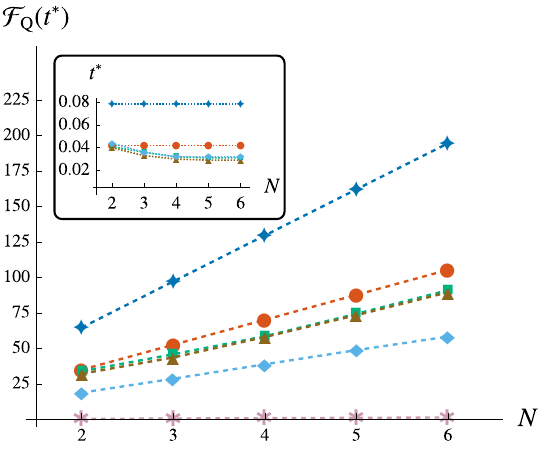}
    \label{subfig:scaling_high_bath_temperature}} 
    \\
    \subfloat[$\beta=0.5$]{\includegraphics[width=0.87\linewidth]{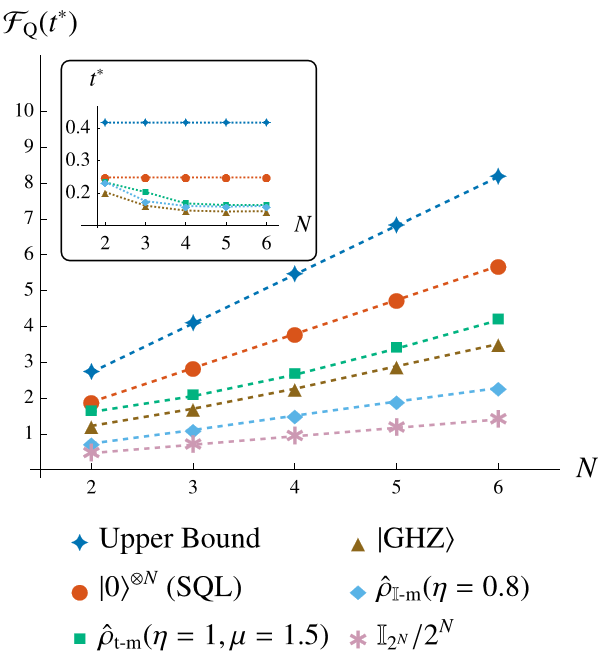}
    \label{subfig:scaling_low_bath_temperature}}
    \caption{\justifying  Maximal QFI over time as a function of the number $N$ of thermometers. Different colors and line styles correspond to the initial states of the thermometers shown in the legend. The ground state (red dots) achieves the best performance and saturates the SQL, while the maximally mixed state (pink asterisks) performs worst as it shows no transient enhancement. Quantum correlations provide an improvement in the maximal QFI and lead to slight deviations from linear behavior for small $N$.}
    \label{fig:qfi_vs_N}
\end{figure}

As discussed in Sec.~\ref{sec:qfi_bound}, no superlinear scaling of the QFI with the number $N$ of thermometers is achievable by any input state at any fixed time of the dynamics, if the encoding of $\beta$ is performed via the application of $N$ independent GAD quantum channels. In spite of this constraint, we examine here how the time-optimized QFI varies with the number of thermometers for different input states, while keeping the bath temperature fixed at a value that generally permits the emergence of a transient QFI peak.

In Fig.~\ref{fig:qfi_vs_N}, we plot the time-optimized QFI obtained from the following initial states: $\ket{0}^{\otimes N}$, $\thmix(\eta=1,\mu=1.5)$, $\ket{\text{GHZ}}$, $\idmix(\eta=0.8)$, and $\id_{2^N}/2^N$, with the number of qubits taking values in the set $N\in\{2,3,4,5,6\}$. For the inverse bath temperature, we use: $\beta= \{0.1,0.5\}$. We denote by $t^*$ the time instant at which the QFI of the evolved $N$-qubit state attains its maximal value. The figure also displays the (linear) scaling of the time-optimized upper bound in Eq.~\eqref{eq:bound}. In the two insets of Fig.~\ref{fig:qfi_vs_N}, we show $t^*$ as a function of $N$ for the different input states; we also include the time instants that maximize the upper bound~\cite{qui_quo_e_pure_qua}.

By initializing the thermometers' ensemble in the product states $\ket{0}^{\otimes N}$ and $\id_{2^N}/2^N$, the scaling of the time-optimized QFI is strictly linear due to the additivity property of the QFI for separable states. Additionally, the optimal time to extract information from the thermometers is the same for all $N$. Indeed, employing an $N$-qubit product state as input state is formally equivalent to conducting $N$ independent single-qubit experiments.

For the initially correlated $\thmix$, $\ket{\text{GHZ}}$ and $\idmix$ states, we observe deviations from a perfectly linear trend at small $N$. However, as the number of thermometers increases, the QFI approaches a linear scaling, characterized by a specific slope that must hold up to the large-$N$ limit (due to the upper bound in Eq.~\eqref{eq:bound}). Moreover, the time at which the QFI is maximized varies for these states: it decreases with increasing system size, approaching a constant value for $N>4$.

We observe that the upper bound is never saturated. Among all the states considered, and for each value of $N$, the product state $\ket{0}^{\otimes N}$, obtained from the ground state of the Hamiltonian in Eq.~\eqref{eq:hamiltonian}, yields the largest QFI value. The corresponding curve also determines the SQL, as given in Eq.~\eqref{SQL}. The gap between the upper bound and the SQL increases as the bath temperature increases, as can be inferred from the comparison between Figs. \ref{subfig:scaling_high_bath_temperature} and \ref{subfig:scaling_low_bath_temperature}. Furthermore, by comparing the two panels, we observe that the difference between the QFI values obtained with (i) the highly entangled states $\thmix(\eta=1,\mu=1.5)$ and $\ket{\text{GHZ}}$, and (ii) the ground state decreases as the bath temperature increases. This is consistent with the results obtained in Sec.~\ref{sec:intermplay_correlations_localtemperature}, Fig.~\ref{fig:Gcorr_thmix_vs_beta}, for small values of $\beta$.

The poorest performance is always provided by the maximally mixed state. Since the QFI associated with this state never displays any peaks during the transient regime, its maximum is always attained at thermal equilibrium. Therefore, the corresponding curve is accurately described by Eq. \eqref{eq:QFI_thermal} and can serve as a reference for the precision achievable in equilibrium thermometry. Comparisons with the other curves in Fig.~\ref{fig:qfi_vs_N} further highlight the precision advantage offered by the nonequilibrium approach. The discrepancy between early-stage and asymptotic temperature estimation becomes more pronounced at higher bath temperatures (corresponding to smaller~$\beta$), as shown in Fig.~\ref{subfig:scaling_high_bath_temperature}, owing to the amplification of the transient peaks of the QFI. 

\section{Conclusions}

In this paper, we analyze the fundamental limits of temperature estimation for a macroscopic thermal bath using an ensemble of initially correlated qubit-thermometers undergoing weak, Markovian interactions with the bath. By tracking the time evolution of the QFI during the thermalization process, our study confirms that quantum thermometry performed in the transient regime offers a significant advantage over thermometry at equilibrium, in agreement with the literature on the subject~\cite{PhysRevE.110.024132,Vasco2018,Fraz_o_2024, alves_landi_2022, marti_paper}. This enhancement arises from the ability to harness resources endowed in the input thermometers' state, including quantum correlations. The analysis further characterizes how the QFI depends on the bath temperature and on local properties of each thermometer, such as coherence and population terms. Indeed, we have verified that lower the local temperature and higher the amount of quantum coherence of the initial thermometers' state, better the thermometric performance. This observation is fully consistent with the existing literature.

Our paper presents several key results that constitute its main novelty. First, we show that quantum correlations in the initial state of the thermometers' ensemble lead to an enhancement of the nonequilibrium thermometry precision, if compared to the performance obtained when initializing the thermometers in the tensor product of their single-qubit reduced states. 
In other words, correlated states outperform uncorrelated ones with same local finite-temperature and coherence. Second, we illustrate that highly correlated input states---despite featuring high local temperatures and/or lacking local coherence---can still surpass the performances of separable states with lower local temperature, saturating the sensitivity of the optimal state $\ket{0}^{\otimes N}$ in the limit of extremely hot thermal baths.

None of the tested entangled input states surpasses the sensitivity of the $\ket{0}^{\otimes N}$ state, corresponding to the SQL. However, the gap with the upper bound on the QFI allows for the existence of a quantum correlated input state capable of achieving a precision beyond the SQL. On top on that, the ground state is thermodynamically challenging to prepare, according to the third law of thermodynamics~\cite{BuffoniPRL2022}, as it corresponds to a zero-temperature pure state.  This is confirmed by the fact that many studies are currently being conducted on high-fidelity preparation of low-temperature states in specific, well-controlled laboratory settings~\cite{Buffoni2023cooperativequantum,Buffoni_2025}; the problem is of obvious relevance for quantum computing~\cite{PhysRevLett.113.220501}. Still, initializing the ensemble of thermometers in a quantum-correlated state with a local temperature significantly greater than zero appears quite more feasible.

We have identified some main outlooks of this work. 
(i) Firstly, it would be interesting to optimize the channel QFI on the pure state space for small values of $N$. This would allow us to uniquely identify the state with the best performance and examine its defining characteristics. 
(ii) As emphasized in the discussion, the peak of the QFI for entangled input states, although less pronounced, occurs earlier than for the maximally locally-coherent product state $\ket{+}^{\otimes N}$. This may provide an experimental advantage in practical scenarios where the accessible experimental time is limited by external disturbances. Indeed, achieving enhanced precision at shorter times reduces the overall duration of the experiment and, consequently, the impact of unwanted interactions with external systems that could degrade the encoding of the bath temperature in the thermometers' state. (iii) The third would be to design a protocol
achieving a global thermometric analysis~\cite{RubioPRL2021,boeyens_2021}, whereby no substantial prior knowledge on the bath temperature is available to us. As a matter of fact, the precision enhancement in nonequilibrium thermometry can only be achieved within a narrow time window, whose position depends on the initially unknown value of $\beta$. 
A step towards this goal could be made by introducing an initial equilibrium experiment, which would supply prior information about the bath temperature. Then, this estimate could be used to approximately identify the time interval where the QFI reaches its nonequilibrium peak.
(iv) Furthermore, although entangled initial states would not guarantee optimal sensitivity in terms of QFI, they may prove to be more effective in enhancing the (classical) Fisher information, when one is constrained to use an experimentally-feasible, non-optimal estimation basis instead of the optimal one. (v) It would also be interesting to design a quantum channel for performing nonequilibrium thermometry such that the upper bound allows for a superlinear scaling of the QFI with $N$ (see Appendix~\ref{sec:m2_is_null} for further detail). It would be already quite interesting to identify a channel with these characteristics even within a short time window, and verify if the Heisenberg scaling can be observed in the nonequilibrium regime. In this regard, a promising direction would be to consider nonlocal maps, which introduce quantum correlations among the thermometers over time, while they undergo thermalization. In this regard, the Authors of Ref.~\cite{marti_paper} have recently demonstrated that a quadratic scaling of the QFI with the number of thermometers can be achieved through continuous measurement and re-initialization of the thermometers in their ground state, while they undergo collective dynamics. Hence, proceeding along this line, it would worth determining the minimal conditions that allows for a quadratic scaling of the QFI when quantum coherence and entanglement are present in the initial state of the thermometers or generated during their collective (non-local) thermalization. 

\section*{Data Availability Statement}

The data plotted in the Figs.~\ref{fig:effect_of_beta}-\ref{fig:qfi_vs_N} are available at~\cite{EnricoGithub}.

\section*{Acknowledgments}

The authors acknowledge very useful discussions with Lorenzo Buffoni, Angelo Carollo, Beatrice Donelli, Gonçalo Frazao, Seth Lloyd, Yasir Mehmood, and Andrea Smirne. 
This work has been financial supported by the PNRR MUR project PE0000023 NQSTI funded by the European Union---Next Generation EU, the PRIN project 2022FEXLYB ``Quantum Reservoir Computing (QuReCo)'', and the European project ``High Performance Computer and Quantum Simulator hybrid (HPCQS)'' with grant agreement No. 101018180 funded by the European Union's Horizon 2020 research and innovation programme.

\appendix
 
\section{Lindblad formulation of the GAD channel dynamics}
\label{sec:appendix_master_equation}

We model the local Markovian map evolving the state of each thermometer qubit via a GAD quantum channel. In addition to what discussed in the main text, we show that this dynamics can be also described by means of the Gorini-Kossakowski-Lindblad-Sudarshan quantum master equation
\begin{equation}
    \frac{\partial \hat{\rho}}{\partial t} = - \frac{ i }{\hbar} [\hat{H},\hat{\rho}] + \sum_{j \neq k} \left( \hat{\mathcal{L}}_{jk} \hat{\rho} \hat{\mathcal{L}}_{jk}^\dagger - \frac{1}{2} \left\{ \hat{\mathcal{L}}_{jk}^\dagger \hat{\mathcal{L}}_{jk}, \hat{\rho} \right\} \right),
    \label{eq:lindblad}
\end{equation}
where $\hat{H}$ is the qubit Hamiltonian in Eq.~\eqref{eq:hamiltonian} and $\hat{\mathcal{L}}_{ij}$ are the jump operators responsible for the Markovian thermalization dynamics, which read as
\begin{equation}\label{eq:jump_operators}
    \hat{\mathcal{L}}_{jk} = \sqrt{\Gamma_{jk}} |j\rangle\langle k|,
\qquad j, k \in \{0,1\}.
\end{equation}
In Eq.~\eqref{eq:jump_operators}, 
\begin{equation}
    \Gamma_{jk} =
\begin{cases}
        \gamma (n_{jk} + 1) \quad &j<k \\
        0 \quad &j=k \\
        \gamma n_{jk} \quad &j>k
\end{cases}
\end{equation}
are the transition rates from state $k$ to state $j$, with $\gamma>0$ denoting the evolution rate of the single thermometer, while $n_{jk}$ are the thermal ratios $n_{jk} = 1/( e^{\beta (\varepsilon_k - \varepsilon_j)} - 1 )$.

The solution of the quantum master equation in Eq.~\eqref{eq:lindblad} evolves the single-qubit density operator with entries $\rho_{jk}$ into the state 
\begin{equation}\label{eq:gad_phi}
\hat{\rho}_{\rm out}^{(1)}(t,\beta) 
= \begin{pmatrix}
        e^{-\lambda t} (\rho_{11}- \pi_0 ) & e^{-\frac{\lambda}{2}t - it} \rho_{12}  \\
        e^{-\frac{\lambda}{2}t + it} \rho_{21} &   e^{-\lambda t} (\rho_{22} - \pi_1)
    \end{pmatrix}
    + \hat{\rho}_{\rm th}(\beta),
\end{equation}
with $\pi_1 = 1 - \pi_0$. Eqs.~\eqref{eq:gad_kraus} and \eqref{eq:gad_phi} provide the same quantum state apart from the off-diagonal phase $e^{-it}$ stemming from the coherent dynamics $ i [\hat{H},\hat{\rho}]/\hbar$ in Eq.~\eqref{eq:lindblad}. This similarity allows us to determine the physical meaning of the parameters $p, q$ entering the Kraus operators of Eq.~\eqref{eq:gad_kraus}. If we neglect the oscillations in the off-diagonal terms in Eq.~\eqref{eq:gad_phi}, the thermalization process can be described by the GAD channel.

From Eqs.~\eqref{eq:gad_kraus} and~\eqref{eq:gad_phi}, it follows that the evolution of the populations is unaffected by the initial coherences, and similarly, the evolution of the coherences is independent of the initial populations. The exponential factors governing the thermalization of the diagonal and off-diagonal elements of each thermometer's state differ by a factor $1/2$.

\section{Details on the QFI at thermal equilibrium}
\label{sec:qfi_thermal}

It is known~\cite{Vasco2018} that for diagonal states---and thus, in particular, for asymptotic thermal states---the relation $\mathcal{F}_Q (\hat{\rho}_{\text{th}}(\beta))=\Delta^2\hat{H}$ holds. We can therefore evaluate the asymptotic value of the QFI from the variance of the thermometer Hamiltonian w.r.t.~the thermal state at the temperature $T$ (inverse temperature $\beta=1/T$) under consideration. Following Ref.~\cite{Fraz_o_2024}, we have
\begin{equation}
    \langle \hat{H}\rangle = \varepsilon_0 \pi_0 (\beta)  + \varepsilon_1 \pi_1 (\beta)  = \varepsilon_1 - [\varepsilon_1 - \varepsilon_0] \pi_0 (\beta)
\end{equation}
and therefore
\begin{eqnarray}
    \var{ \hat{H} } 
    &=& \left( \langle \hat{H}\rangle - \varepsilon_0 \right)^2 \pi_0 (\beta)
        + \left( \langle \hat{H}\rangle - \varepsilon_1 \right)^2 \pi_1 (\beta) = \nonumber\\
    &=& \Big[ (\varepsilon_1 - \varepsilon_0)[1-\pi_0 (\beta)] \Big]^2 \pi_0(\beta) \nonumber \\
    &&+ \Big[(\varepsilon_1 - \varepsilon_0)  \pi_0 (\beta) \Big]^2 [1- \pi_0(\beta)] = \nonumber \\
    &=& (\varepsilon_1 - \varepsilon_0)^2 \pi_0 (\beta) [1-\pi_0(\beta)].
\end{eqnarray}
One can see that the maximum value of the equilibrium QFI is attained for $\pi_0 = \pi_1 = 1/2$, corresponding to the limit of infinite bath temperature, $T \to +\infty$. In contrast, the QFI vanishes in the zero-temperature limit, $T \to 0$. This behavior is consistent with that of the non-equilibrium QFI as a function of the bath temperature, as discussed in Sec.~\ref{sec:temperature_role}. Therefore, the limit on the estimation precision for $\nu$ repetition of the experiment given by the quantum Cram\'{e}r-Rao bound becomes
\begin{equation}
    \Delta \tilde{\beta } \ge \frac{1}{\hbar \omega\sqrt{\nu \pi_0(\beta)[1- \pi_0(\beta)] }},
\end{equation}
where we substituted the values for the energy eigenvalues $\varepsilon_{0(1)} = -(+) \hbar \omega/2$.

\section{Proof of Eq.~\eqref{eq:bound}}
\label{sec:m2_is_null}

In Eq.~\eqref{channel_qfi}, we have defined the channel QFI of the channel $(\Phi_\beta^{[t]})^{\otimes N}$, denoted by $\mathcal{F}_Q(\Phi_\beta^{[t]})$. The channel acts independently on each qubit probe; therefore, the channel QFI can be upper-bounded as~\cite{Fujiwara_2008, Demkowicz_Dobrza_2013}: 
\begin{equation}
\mathcal{F}_Q\left(\Phi_\beta^{[t]}\right) \leq 4 \min_{\mathcal{K}'} \left[ N \norm{\hat{A}_{\mathcal{K}'}} + N(N-1) \norm{\hat{B}_{\mathcal{K}'}}^2\right], \label{up_bound}
\end{equation}
where
\begin{equation}
\hat{A}_{\mathcal{K}'} = \sum_{j=1}^4 \frac{\partial \hat{K}'_j{}^\dagger}{\partial \beta} \frac{\partial \hat{K}'_j}{\partial\beta},
\qquad
\hat{B}_{\mathcal{K}'} = i \sum_{j=1}^4 \frac{\partial \hat{K}_j'{}^\dagger}{\partial \beta} \hat{K}'_{j}.
    \label{eq:A_and_B}
\end{equation}
In these equations, $\mathcal{K}'$ denotes an equivalent Kraus representation of the GAD channel, obtained from the Kraus operators in Eq.~\eqref{eq:GAD_Kraus_operators} by means of $\hat{K}'_j=\sum_{l=1}^4 U_{jl}(\beta)\hat{K}_l \ (j=1,\dots,4)$, where $U_{jl}(\beta)$ is a $4\times4$ unitary matrix that depends on $\beta$. A modified version of the bound in Eq.~\eqref{up_bound} holds also for unitary encoding channels that are affected by noise, even when the measurement on the probes' state is performed adaptively~\cite{kurdzialek2023using}.  

According to Eq.~\eqref{up_bound}, finding a single representation $\mathcal{K}'$ such that $\lVert\hat{B}_{\mathcal{K}'}\rVert=0$ constitutes a sufficient condition to establish the asymptotic linear scaling of $\mathcal{F}_Q(\Phi^{[t]}_{\beta})$ with $N$, the total number of qubits. In our case, one can actually show that the original Kraus representation already provides a valid option for obtaining this scaling. Indeed, by evaluating the derivatives of the Kraus operators in Eq.~\eqref{eq:GAD_Kraus_operators}, we get
\begin{eqnarray}
    \partial_\beta \hat{K}_1^\dagger &=& 
    \begin{pmatrix}
        \frac{\partial_\beta q}{2 \sqrt{q}} & 0 \\
        0 &-  \frac{\partial_\beta p }{2 \sqrt{1-p}}\sqrt{q} + \frac{\partial_\beta q}{2 \sqrt{q}} \sqrt{1-p}
    \end{pmatrix},\nonumber \\
     \partial_\beta \hat{K}_2^\dagger &=& 
    \begin{pmatrix}
    0 & 0 \\
    \frac{\partial_\beta p}{2 \sqrt{p}}\sqrt{q} + \frac{\partial_\beta q}{2 \sqrt{q}}\sqrt{p} & 0
    \end{pmatrix},\nonumber \\
     \partial_\beta \hat{K}_3^\dagger &=& 
     \begin{pmatrix}
         - \frac{\partial_\beta p}{2 \sqrt{1-p}} \sqrt{1-q} - \frac{\partial_\beta q}{2 \sqrt{1-q}} \sqrt{1-p} & 0 \\
         0 & - \frac{\partial_\beta q}{2 \sqrt{1-q}}
     \end{pmatrix},\nonumber \\
     \partial_\beta \hat{K}_4^\dagger &=& 
     \begin{pmatrix}
         0 &   \frac{\partial_\beta p}{2 \sqrt{p}} \sqrt{1-q} - \frac{\partial_\beta q}{2 \sqrt{1-q}} \sqrt{p} \\
         0 & 0
     \end{pmatrix}.
\end{eqnarray}
By then multiplying each derivative by the corresponding Kraus operator
\begin{eqnarray}
    \partial_\beta (\hat{K}_1^\dagger)  \hat{K}_1 &&= 
    \frac{1}{2} \partial_\beta
    \begin{pmatrix}
        q & 0 \\
        0 & q (1-p)
    \end{pmatrix},\nonumber \\
    \partial_\beta (\hat{K}_2^\dagger)  \hat{K}_2 &&= 
     \frac{1}{2} \partial_\beta
    \begin{pmatrix}
        0 & 0 \\
        0 & pq
    \end{pmatrix},\nonumber \\
    \partial_\beta (\hat{K}_3^\dagger)  \hat{K}_3 &&=
     \frac{1}{2} \partial_\beta
    \begin{pmatrix}
        (1-p)(1-q) & 0 \\
        0 & -q
    \end{pmatrix},\nonumber \\
    \partial_\beta (\hat{K}_4^\dagger)  \hat{K}_4 &&= 
     \frac{1}{2} \partial_\beta
    \begin{pmatrix}
    p(1-q) & 0 \\
    0 & 0
    \end{pmatrix},
\end{eqnarray}
and taking the sum of the resulting terms, we find that $\hat{B}_{\mathcal{K}}$ vanishes identically. Notably, as demonstrated by the foregoing calculation, this result holds for arbitrary functional forms of the coefficients $q(\beta)$ and $p(\beta)$.

Few additional steps are now required to complete the proof of Eq.~\eqref{eq:bound}. We notice that both $\hat{A}_{\mathcal{K}'}$ and $\hat{B}_{\mathcal{K}'}$ depend on $\beta$ only through $\mathcal{K}'$ and its first derivative with respect to $\beta$. Moreover, they are insensitive to changing the Kraus representation with a $\beta-$independent unitary transformation. Therefore, in view of the optimization prescribed by Eq.~\eqref{up_bound}, the parametrization $U_{jl}(\beta)=e^{-i\beta\hat{h}}$, where $\hat{h}$ is a $4\times4$ hermitian matrix, appears to be sufficiently general. The optimization can thus be equivalently expressed in terms of $\hat{h}$. Moreover, with this parametrization we have $\partial\hat{K}_j'/\partial\beta=\partial\hat{K}_j/\partial\beta-i\sum_{l=1}^4h_{jl}\hat{K}_l$. Being primarily interested in the asymptotic linear scaling of the channel QFI with $N$, we derive a further upper bound on $\mathcal{F}_Q(\Phi_{\beta}^{[t]})$:
\begin{align}
    \mathcal{F}_Q\left(\Phi_\beta^{[t]}\right) &\leq 4 \min_{\hat{h}} \left[ N \norm{\hat{A}_{\mathcal{K}'}} + N(N-1) \norm{\hat{B}_{\mathcal{K}'}}^2\right]\nonumber\\
    &\leq 4N \min_{\hat{h}} \norm{\hat{A}_{\mathcal{K}'}}.
\end{align}
Note that, unlike the optimization over $\hat{h}$ appearing in the first line, the optimization in the second line, which corresponds to Eq.~\eqref{eq:bound}, is constrained and subject to the condition
\begin{equation}
    \sum_{jl=1}^4 h_{jl}\hat{K}_j^{\dagger}\hat{K}_l=i\sum_{j=1}^4 \frac{\partial \hat{K}_j^\dagger}{\partial \beta} \hat{K}_{j},
\end{equation}
where the Kraus operators are those of Eq.~\eqref{eq:GAD_Kraus_operators}. This linear constraint on $\hat{h}$ comes from restricting the optimization to those hermitian matrices that preserve the condition $\hat{B}_{\mathcal{K}'}=0$. As proven in Ref.~\cite{Demkowicz_Dobrza_ski_2012}, this constrained minimization problem can be reformulated as a semi-definite program. This approach has been employed for our numerical computation of the upper bound. 

\section{Proof of Eq.~\eqref{eq:reduced_plusghz_state}}
\label{sec:reduced_plusghz}

We first introduce the unnormalized +-s state as defined in Eq.~\eqref{eq:k_GHZ_superposition}
\begin{equation}
    \ket{\tilde{\psi}_\text{+-s}} = \sin \alpha \ket{\text{GHZ}} + \cos\alpha \ket{+}^{\otimes N}
\end{equation}
We observe that the tensor product of $N$ $\ket{+}$ states can be rewritten as
\begin{equation}
    \ket{+}^{\otimes N} = \frac{1}{2^{\frac{N}{2}}}\sum_{x=0}^{2^N-1} \ket{\overline{x}}
\end{equation}
where the overlined state $\ket{\overline{x}}$ is the vector of an $N$-qubit state defined by the binary representation of $x$. Therefore, we have
\begin{eqnarray}
    \ket{\tilde{\psi}_\text{+-s}} &=& \left( \frac{\sin(\alpha)}{\sqrt{2}} + \frac{\cos(\alpha)}{2^{\frac{N}{2}}} \right)\left( \ket{0}^{\otimes N} + \ket{1}^{\otimes N} \right) +\nonumber \\
    &+&\frac{\cos(\alpha)}{2^{\frac{N}{2}}} \sum_{x=1}^{2^N-2} \ket{\overline{x}}
\end{eqnarray}
Now we compute the inner product of $\ket{\tilde{\psi}_\text{+-s}}$ with itself, to obtain the normalization constant of the state. As all vectors on the RHS are orthonormal,
\begin{align}
\braket{\tilde{\psi}_\text{+-s} | \tilde{\psi}_\text{+-s} } &= 2 \left( \frac{\sin(\alpha)}{\sqrt{2}} + \frac{\cos(\alpha)}{2^{\frac{N}{2}}} \right)^2 + \frac{2^{N}-2}{2^N} \cos^{2}(\alpha) = \notag \\
    &= 1 + 2^{\frac{1-N}{2}} \sin (2 \alpha) = C(+,\alpha,N)^{2}.
\end{align}

Then, we evaluate the unnormalized density operator for the $+$-s N-qubit state
\begin{align}
    & \ket{\tilde{\psi}_\text{+-s} }\bra{\tilde{\psi}_\text{+-s}} = \notag \\
    & = \left( \frac{\sin(\alpha)}{\sqrt{2}} + \frac{\cos(\alpha)}{2^{\frac{N}{2}}} \right)^{2} \big(\ket{0}^{\otimes N} + \ket{1}^{\otimes N} \big) \big(\bra{0}^{\otimes N} + \bra{1}^{\otimes N} \big) \notag \\
    & + \left( \frac{\sin(\alpha)}{\sqrt{2}} + \frac{\cos(\alpha)}{2^{\frac{N}{2}}} \right) \frac{\cos(\alpha)}{2^{\frac{N}{2}}} \bigg[ \big(\ket{0}^{\otimes N} + \ket{1}^{\otimes N} \big) \times \notag \\ 
    &\times \sum_{x=1}^{2^N-2} \bra{\overline{x}} + {\rm c.c} \bigg] 
    + \frac{\cos^{2}(\alpha)}{2^N} \sum_{x,y=1}^{2^N-2} \ket{\overline{x}}\bra{\overline{y}}.
\label{eq:rho_tildepsi_plusghz}
\end{align}
Now we can evaluate the reduced density operator of the first qubit, considering the following expression
\begin{equation}
    \sum_{x=1}^{2^N-2} \ket{\overline{x}} = \ket{0} \sum_{m=1}^{2^{N-1}-1} \ket{\overline{m}} + \ket{1} \sum_{m=0}^{2^{N-1}-2} \ket{\overline{m}} .
\end{equation}

Only the first and the third terms of Eq.~\eqref{eq:rho_tildepsi_plusghz} give a contribution to the populations of the reduced density operator. Notably, the populations turn out to be equal, namely
\begin{align}
    \big[\tr_{N-1}[\tilde{\rho}_\text{+-s}]\big]_{11} & = \left( \frac{\sin(\alpha)}{\sqrt{2}} + \frac{\cos(\alpha)}{2^{\frac{N}{2}}} \right) ^2 + \frac{2^{N-1}}{2^N} \cos^{2}(\alpha) = \notag \\
    & = \frac{1}{2} C(+,N,\alpha)^2 = [\tilde{\rho}_\text{+-s}]_{22}.
\end{align}
Therefore $[\rho_\text{+-s}]_{11} = [\rho_\text{+-s}]_{22} =1/2$. The elements outside the diagonal Only the second or third terms contribute to the coherence term
\begin{align}
    \big[\tr_{N-1}[\tilde{\rho}_\text{+-s}]\big]_{12} & = 2 \left( \frac{\sin (\alpha)}{\sqrt{2}} + \frac{\cos(\alpha)}{2^{\frac{N}{2}}} \right) \frac{\cos(\alpha)}{2^{\frac{N}{2}}} + \notag \\
    & \hspace{2cm} + \frac{2^{N-1}-2}{2^N} \cos^{2}(\alpha) = \notag \\
    &= \frac{1}{2} \left( 2^{\frac{1-N}{2}} \sin (2 \alpha) + \cos^{2}(\alpha) \right).
\end{align}
Dividing for the squared normalization constant $C(+,N,\alpha)^2$ gives Eq.~\eqref{eq:reduced_plusghz_state}.

\section{Proof of Eq.~(\ref{trace_squeezed})}
\label{sec:appendix_trace_squeezed}

In order to derive the expression of the partial trace given in Eq.~(\ref{trace_squeezed}), let us analyze the squeezed spin state of Eq.~(\ref{eq:squeezed_state}). Recall that this state belongs to the bosonic subspace of the $N$-qubit Hilbert space, and can therefore be expressed as a combination of $N+1$ orthogonal Fock states:
\begin{equation}
    \ket{\psi_{\rm sq} (\chi)}=\frac{1}{2^{N/2}}\sum_{k=0}^N \sqrt{\binom{N}{k}} e^{-i\chi\left(\frac{N}{2}-k\right)^2}|N-k,k\rangle. 
\end{equation}
Here, the notation $|N-k,k\rangle$ refers to a bosonic Fock state with $N-k$ particles in the qubit state $|0\rangle$ and $k$ particles in the qubit state $|1\rangle$ and $\binom{N}{k}$ denotes the binomial coefficient $N!/[(N-k)!k!]$.

The density operator associated with the squeezed state can thus be expressed as 
\begin{align}\label{density_squeezed_explicit}
    \squeez (\chi)=&\frac{1}{2^N}\sum_{k,k'=0}^N\sqrt{\binom{N}{k}\binom{N}{k'}}e^{i\chi\left(N-k-k'\right)\left(k-k'\right)}\times\nonumber\\
    &\qquad\times|N-k,k\rangle\langle N-k',k'|.
\end{align}
The partial trace $\hat{\rho}^{(1)}_{\text{sq}}(\chi)=\tr_{N-1} \left[ \hat{\rho}_{\text{sq}}(\chi)\right]$ can now be evaluated by tracing over any subset of $N-1$ qubits, since the state is invariant under particle exchange. To show how this calculation is carried out, we rewrite the Fock state $\ket{N-k,k}$ in a more explicit form:
\begin{equation}
    \ket{N-k,k}=\frac{1}{\sqrt{M_{N,k}}}\sum_{P \in S_N} \ket{\psi_{P_1}}\otimes\dots\otimes\ket{\psi_{P_N}}.
\end{equation}
In this equation, $P$ is a permutation in $S_N$, the symmetric group of order $N$, $M_{N,k}=N!(N-k)!k!$ is a normalization constant and $\ket{\psi_{P_j}}\in \{\ket{0},\ket{1}\}$ so that $\ket{0}$ and $\ket{1}$ appear $N-k$ and $k$ times, respectively, in the tensor product. Using the above equation, we obtain
\begin{align}\label{partial_Dicke}
&\tr_{N-1}\big[ |N-k,k\rangle\langle N-k',k'| \big]=\frac{1}{\sqrt{M_{N,k}\ M_{N,k'}}}\times \nonumber\\
&\quad \times \sum_{P,P'\in S_N}\ket{\psi_{P_1}}\bra{\psi_{P'_1}}\prod_{j=2}^N\langle \psi_{P_j}|\psi_{P'_j}\rangle,
\end{align}
where, without loss of generality, we chose to trace over the last $N-1$ qubits. Due to the $N-1$ scalar products that appear in this formula, the partial trace will vanish for all pairs $k,k'$ that do not fulfill the condition $k-k'=0,\pm 1$. The combination $k-k'=0$ will provide terms proportional to $\ket{0}\bra{0}$ and $\ket{1}\bra{1}$, while $k-k'=-1$ and $k-k'=1$ terms proportional to $\ket{0}\bra{1}$ and $\ket{1}\bra{0}$, respectively. Through simple combinatorics, it is possible to derive explicit expressions for all the relevant partial traces in Eq.~(\ref{partial_Dicke}). By combining these with Eq.~\ref{density_squeezed_explicit}, we then find
\begin{align}
    \bra{0}\hat{\rho}^{(1)}_{\text{sq}}(\chi)\ket{0}&=\frac{1}{2^N}\sum_{k=0}^{N-1}\binom{N-1}{k}=\frac{1}{2},\\
    \bra{1}\hat{\rho}^{(1)}_{\text{sq}}(\chi)\ket{1}&=\frac{1}{2^N}\sum_{k=1}^N \binom{N-1}{k-1}=\frac{1}{2},\\
    \bra{0}\hat{\rho}^{(1)}_{\text{sq}}(\chi)\ket{1}&=\frac{1}{2^N}\sum_{k=0}^{N-1}\binom{N-1}{k}e^{-i\chi\left(N-2k-1\right)}=\nonumber\\
    &=\frac{\cos^{N-1}(\chi)}{2},\\
    \bra{1}\hat{\rho}^{(1)}_{\text{sq}}(\chi)\ket{0}&=\frac{1}{2^N}\sum_{k=1}^{N}\binom{N-1}{k-1}e^{i\chi\left(N-2k+1\right)}=\nonumber\\
    &=\frac{\cos^{N-1}(\chi)}{2}.
\end{align}
We have thus obtained the elements of the single-qubit reduced state given in Eq.~(\ref{trace_squeezed}).

\bibliography{updated_bibliography_1}

\begin{thebibliography}{78}
\bibitem [{plu()}]{pluto}%
  \BibitemOpen
  \href@noop {} {}\bibinfo {note} {A more formal way to write the quantum
  Cram\'{e}r-Rao bound involves the standard deviation $\Delta \tilde{\beta}$,
  where $\tilde{\beta}$ is an estimator of $\beta$. An estimator is any
  function that associates an estimate of $\beta$ with every possible set of
  outcomes obtained from measurements performed on the output state. Moreover,
  the form of the quantum Cram\'{e}r-Rao bound holds only for unbiased
  estimators, which satisfy $\langle \tilde{\beta}\rangle -
  \beta=0$.}\BibitemShut {Stop}%
\bibitem [{top()}]{topolino}%
  \BibitemOpen
  \href@noop {} {}\bibinfo {note} {Absence of quantum correlation defines the
  set of separable states: $\hat{\rho}_{\rm sep}=\sum_{q}p_q
  \hat{\rho}_q^{(1)}\otimes\dots\otimes\hat{\rho}_q^{(N)}$, where $\sum_q
  p_q=1$. Due to the convexity of the QFI and to its additivity on product
  states, we have $\mathcal{F}_Q((\Phi_{\beta}^{(t)})^{\otimes
  N}\hat{\rho}_{\rm sep}) \leq \sum_q
  p_q\sum_{j=1}^N\mathcal{F}_Q(\Phi_{\beta}^{(t)}\hat{\rho}_q^{(j)}) \leq N
  \max_{\rho^{(1)}}\mathcal{F}_Q(\Phi_{\beta}^{(t)}\hat{\rho}^{(1)})$. Since
  the last term in the inequality chain corresponds to the QFI of a product
  state, which belongs to the set of separable states, we actually have
  $\max_{\hat{\rho}_{\rm sep}}\mathcal{F}_Q((\Phi_{\beta}^{(t)})^{\otimes
  N}\hat{\rho}_{\rm sep})=N
  \max_{\rho^{(1)}}\mathcal{F}_Q(\Phi_{\beta}^{(t)}\hat{\rho}^{(1)})$. This
  justifies the definition of the Standard Quantum Limit given in the main
  text.}\BibitemShut {Stop}%
\bibitem [{not()}]{note_wolfram}%
  \BibitemOpen
  \href@noop {} {}\bibinfo {note} {These evaluations are primarily carried out
  using the Wolfram Mathematica software \cite{Mathematica}. The state
  evolution is implemented analytically. This approach allows for the
  analytical computation of the derivative of the state with respect to $\beta$
  , required to solve the Lyapunov equation \eqref{eq:lyapunov} and determine
  the Symmetric Logarithmic Derivative (SLD). This step, as well as the final
  numerical evaluation of the QFI, is performed numerically.}\BibitemShut
  {Stop}%
\end{thebibliography}

\end{document}